\documentclass[12pt,letterpaper]{article}
\usepackage[margin=1.2in]{geometry}
\usepackage{setspace}

\usepackage{epsfig, amsthm, amsmath, amssymb, mathrsfs, latexsym, xpatch, xcolor, mathtools, verbatim, hyperref}

\usepackage[most]{tcolorbox}
\usepackage[mathscr]{euscript}
\usepackage[titletoc]{appendix}
\usepackage[normalem]{ulem} 
\usepackage{tabularx}
\usepackage{float} 
\usepackage[margin=1cm, font={small}]{caption} 

\definecolor{cardinal}{rgb}{0.8, 0.0, 0.0}

\newcommand{\op}{\begin{itemize}}
\newcommand{\ed}{\end{itemize}}

\newcommand{\opp}{\begin{quote}}
\newcommand{\edd}{\end{quote}}

\newcommand{\oop}{\begin{tcolorbox}[colback=white]}
\newcommand{\eed}{\end{tcolorbox}}

\newcommand{\ope}{\begin{enumerate}}
\newcommand{\ede}{\end{enumerate}}

\newcommand{\xm}{\item[]}
\newcommand{\im}{\item}

\title{Scientific Realism vs. Anti-Realism: \\Toward a Common Ground}

\author{Hanti Lin \\[0.5em] University of California, Davis \\\texttt{ika@ucdavis.edu}}

\date{December 20, 2024}

\begin{document}

\maketitle

\begin{abstract} \noindent The debate between scientific realism and anti-realism remains at a stalemate, making reconciliation seem hopeless. Yet, important work remains: exploring a common ground, even if only to uncover deeper points of disagreement and, ideally, to benefit both sides of the debate. I propose such a common ground. Specifically, many anti-realists, such as instrumentalists, have yet to seriously engage with Sober's call to justify their preferred version of Ockham's razor through a positive account. Meanwhile, realists face a similar challenge: providing a non-circular explanation of how their version of Ockham's razor connects to truth. The common ground I propose addresses these challenges for both sides; the key is to leverage the idea that everyone values some truths and to draw on insights from scientific fields that study scientific inference---namely, statistics and machine learning. This common ground also isolates a distinctively epistemic root of the irreconcilability in the realism debate.
\\[0.5em] {\bf Keywords}: {\em Scientific Realism, Instrumentalism, Ockham's Razor, Statistics, Machine Learning, Convergence to the Truth.} 
\end{abstract} 

\tableofcontents

\section{An Invitation}

The debate between scientific realism and anti-realism is deep and pervasive; reconciliation seems hopeless. Even so, we must consider how to live with this stalemate. I propose seeking a common ground, even if only to uncover deeper points of disagreement. This opening section outlines an approach to be developed throughout the paper. The motivation is straightforward: since the realism debate partly concerns the epistemology of science, we should explore whether the areas of science that study scientific inference, such as statistics and machine learning, offer valuable lessons. This exploration proves fruitful.

\subsection{The Disagreement}\label{sec-disagreement}

Scientific realists argue that the available evidence justifies belief in the existence of entities like atoms. More broadly, they contend that we are justified in believing in the literal (or approximate) truth of widely accepted claims of mature sciences, especially those concerning unobservable entities. Anti-realists, by contrast, maintain that justified belief should be limited to claims about observable entities or predictions of observable variables.\footnote{See Chakravartty (2017) for a survey on the realism debate.}

This disagreement spans from general epistemological principles to specific cases, the most iconic being Perrin's (1910) experimental work on Brownian motion, which he regarded as strong evidence for the existence of atoms. Philosophers' analyses of Perrin's empirical evidence reveal divergent judgments on the strength of this evidence for atomism. Realists such as Maddy (1997, chap. II.6), Chalmers (2011), and Hudson (2020) find the evidence strong and intuitively compelling, while anti-realists like van Fraassen (2009) disagree. The disagreement deepens when general epistemological principles are considered. For example, realists Achinstein (2001, chap. 12) and Psillos (2011) assess Perrin's evidence using objective epistemic probability, a concept long resisted by subjective Bayesians, including the anti-realist van Fraassen (1989). Thus, the debate encompasses both general principles of evidential support and judgments (intuitive or considered) on specific cases.

The scientific realism debate is unusually fractured compared to other epistemological debates, where shared intuitive judgments often provide some common ground. Consider the debate over external world skepticism: there is broad consensus that beliefs in the existence of some external objects (such as hands) are somehow justified, with disagreements focusing on the underlying principles of justification (Comesa\~{n}a \& Klein 2024, secs. 3 and 4). In contrast, the realism debate features disagreements over both general principles and specific cases, leaving little shared ground.

If resolving such a deep divide is hopeless, the real question is how to live with it. I propose a systematic search for common ground between realists and anti-realists, even if only to isolate core disagreements and uncover deeper, unnoticed issues.

\subsection{A Common Ground Easy to Come By?}

Some might argue that the search for common ground is unnecessary, as one already exists: the method scientists use for theory choice. The term `theory choice', widely used in philosophical discussions of science, is nicely neutral. It suggests that the method shared by the scientific community uses evidence to {\em choose} a theory from alternatives, without prescribing what attitude scientists should adopt toward the chosen theory---whether believing in its approximate truth, trusting its predictions, or something else. Thus, this shared method might appear to provide sufficient common ground, rendering further search unnecessary---or so some might think.

However, such a cheap common ground is an illusion. There is no such thing as {\em the} scientific method of theory choice. Take Ockham's razor, which seeks to balance simplicity with other theoretical virtues, such as goodness of fit with data. Scientists lack consensus on {\em the} right balance. A classic example is the two versions of Ockham's razor, AIC and BIC, developed in the 1970s for statistical model selection. While both favor simplicity, AIC gives it more priority, whereas BIC emphasizes fit with data. These two razors sparked debates that extended beyond statistics to biological and social sciences (Burnham et al. 2011; Aho et al. 2014).

In fact, the history of AIC vs. BIC not only highlights the absence of an easy common ground but also serves to explore the possibility of a solid shared foundation, to which I now turn.

\subsection{Proposal: Seeking a Common Ground}

My initial thought is simple: everyone values some truths. Consider instrumentalists, for instance. The more they emphasize predictive accuracy over truth, the more they should care about:
\opp
{\bf Question Regarding Predictive Accuracy.}
Which model on the table is most predictive?
\edd
\noindent Here is the thing: to such empirical questions, instrumentalists should be interested in seeking {\em true} answers. Realists, by contrast, tend to focus on a different question:
\opp
{\bf Question Regarding Truth.}
Which theory or model on the table is true, if any?
\edd
\noindent Rather than dwelling on the differences between these questions, I propose emphasizing that both sides share a commitment to finding true answers to certain empirical questions. In fact, this shared interest in pursuing truths is not my invention. It emerged during the 1970s and 80s in statistics. Almost the same statistical framework was used to study both the choice of a predictive model (Akaike 1973) and the choice of a true model (Nishii 1984). 

The main idea can be illustrated with an analogy. Data science, encompassing both statistics and machine learning, functions as a {\em consulting service} for scientists as clients. This service evaluates an inference method by examining its capability to find or approximate the truth across a range of possible worlds---the truth being the true answer to {\em whatever} empirical question posed by the client. If the question is which model has the highest predictive power, Akaike's (1973) work applies, recommending the blunter razor AIC (which places less emphasis on simplicity). If instead the question is which model is true, Nishii's (1984) work recommends the sharper razor BIC. Thus, both statisticians Akaike and Nishii operate within the same consulting framework, employing a largely uniform approach to evaluating inference methods. 

This hints at a promising, uniform approach to {\em serious pursuit of truth} for realists and instrumentalists alike---a common ground to be developed throughout this paper. Toward the end, I will show how this common ground helps isolate a particularly deep {\em epistemological} disagreement between the two parties---one that, I will argue, runs deeper than many other issues in the realism debate, such as the {\em semantic} question of interpreting scientific theories or the {\em teleological} question of the aims of science.


\subsection{Benefiting Both Parties}

The common ground I propose is also designed to benefit both parties.

Consider the case of anti-realists. I echo Sober's (2002) call for a positive anti-realist epistemology instead of an over-focus on critiques of realism. This recommendation embodies a philosophical temperament worth striving for. Sober specifically suggests that instrumentalists draw inspiration from Akaike's (1973) development of AIC for {\em predictive} model selection.

While I largely agree with Sober, his and Forster's account of Akaike only tells half the story (Forster \& Sober 1994; Sober 2002). Akaike's work must be understood within a broader statistical tradition that both inspired his contributions and was continued by his successors. I will trace this history and explain why it represents a serious pursuit of truth as described above. For Akaike and his followers, the truth sought is, first and foremost, the true value of each model's predictive power. This instrumentalist pursuit of truth exemplifies a straightforward application of the common ground I propose (section \ref{sec-instrumentalism}). The same observation will be applied to other anti-realist views, such as van Fraassen's (1980) constructive empiricism.

Now consider realists. They have long been challenged to explain how their favored inference methods, such as {\em inference to the best explanation} (IBE), are connected to truth, given their professed commitment to truth-finding. IBE, essentially Ockham's razor on steroids, seeks to balance a broader range of theoretical virtues. Realists often rely on IBE itself to argue for its truth-conduciveness (Putnam 1975). A particularly clear example is Psillos's argument  (1999, 76-77):
\opp 
\begin{center}
{\bf IBE for IBE}
\end{center}

{\sc Premise 1.} The choice of an inference method for prediction in a context of inquiry is often informed by the background theories, which scientists arrive at using IBE.

{\sc Premise 2.} The chosen methods are often reliable for producing true predictions, which require explanation.

{\sc Premise 3.} This predictive reliability is best explained by the reliability of IBE for arriving at approximately true background theories.

{\sc Conclusion.} Therefore, by IBE, IBE is reliable for arriving at approximately true theories.
\edd 
Attempts have been made to explain why such an argument is not viciously circular (Psillos 1999, 79-87). While I do not personally find this particular circle vicious, my concern lies elsewhere. Perhaps it is futile to persuade those already committed to the opposing view, but both realists and anti-realists should still aim to address the undecided. The realists' reliance on this circular reasoning, whether vicious or not, does little to achieve that goal.

I believe that certain untapped resources can explain how realist-friendly inference connects to truth. These resources lie in the common ground I propose, informed by recent advances in formal epistemology and machine learning, originally developed for a specific type of model selection: choosing a true causal Bayes net (Lin 2019; Lin \& Zhang 2020). Drawing on these ideas, I will conduct a case study on Perrin's empirical evidence for atomism. Perrin's realist inference, as I will show, is another version of Ockham's razor, and I will explain---without circularity---how it connects to truth (section \ref{sec-realism}).


Enough with the motivation; it is time to deliver on the promise. I will begin by developing the common ground (section \ref{sec-core}), followed by an explanation of how it benefits both anti-realists (section \ref{sec-instrumentalism}) and realists (section \ref{sec-realism}). Building on this, I will identify a particularly deep disagreement between the two parties (section \ref{sec-deepest}) and conclude with brief comments on the possible roots of the irreconcilability of the realism debate (section \ref{sec-irreconcilability}).

\section{A Little Story of Pursuit of Truth}\label{sec-core}

The phrase `pursuit of truth' is often heard, perhaps so often that it is easy to take it lightly. Yet certain groups of people have shown how seriously it can be taken in scientific inquiry: frequentists in statistics, and learning theorists in computer science and in formal epistemology. These groups have worked separately for too long, making it easy to overlook their shared ideas. Let me, then, tell a unifying story.

\subsection{Preliminaries: The Notion of Empirical Problems}

To begin, I need to clarify what an {\em empirical problem} is, as the framework outlined below evaluates inference methods only within specific contexts---contexts in which an empirical problem is addressed.

\begin{figure}[ht]
	\centering \includegraphics[width=.6\textwidth]{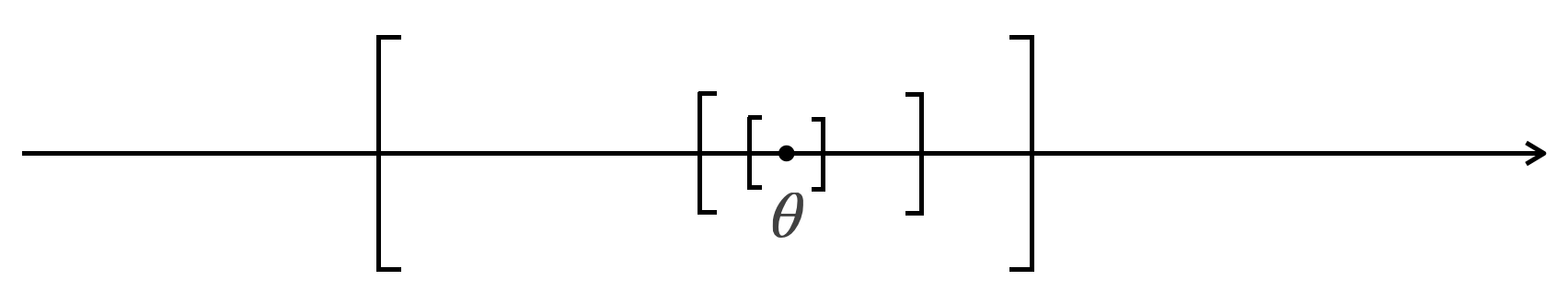}
	\caption{A simple problem of theory choice: `$\theta = 0$' vs. `$\theta \neq 0$'}
	\label{fig-simple-testing}
	\end{figure}

Consider an example. Suppose there is an unknown parameter $\theta$, whose value might be any real number. We pose the following question:
	\opp
	{\bf Question.} Is the true value of $\theta$ exactly 0?
	\edd
An empirical problem consists of not only a {\em question} with potential answers as competing hypotheses but also some {\em background beliefs} or {\em assumptions}, which are essential for evaluating inference methods. For the present problem, let's say we have the following:
	\opp 
	{\bf Background Assumptions.} The parameter $\theta$ may take any real value, including very small ones. Evidence is represented as an interval containing the true (but unknown) value of $\theta$; shorter intervals indicate more evidence. As evidence accumulates, each new interval is nested within the previous one, but the interval never shrinks to a single point.
	\edd 
We can visualize this empirical problem as shown in figure \ref{fig-simple-testing}, where a growing body of evidence is represented by a sequence of progressively shorter intervals.

A quick clarification: This empirical problem is a simplistic example. We could have posed a more interesting question under weaker assumptions, but this simple problem provides a quick introduction to the key ideas I want to convey. 

An inference method for the present problem is a (mathematical) function that, upon receiving a nonzero interval as evidence, outputs one of three options: `$\theta = 0$', `$\theta \neq 0$', or a question mark `\texttt{?}' to indicate suspension of judgment. The key challenge is assessing these methods to determine which are justified. We will first explore how formal learning theorists approach this issue, and then see how frequentist statisticians address it in a similar fashion.

\subsection{Formal Learning Theory}\label{sec-formal-learning}

It would be ideal if there could be an inference method being so good that a prescribed amount of evidence is guaranteed to lead us to the truth. This high standard can be defined as follows:	
\oop  	
	{\bf Definition (Uniform Convergence to the Truth).} An inference method $M$ for an empirical problem is said to meet the standard of {\em uniform convergence to the truth} iff,
		\op 
		\xm there exists a prescribed amount of evidence $n$ (or interval length $\ell$) such that, 
			\op
			\xm in every possible world $w$ compatible with the background assumptions, 
				\op 
				\xm $M$ outputs the answer true at $w$ (`$\theta = 0$' or `$\theta \neq 0$') whenever the evidence reaches or surpasses the prescribed amount $n$ (or the interval length shrinks to $\ell$ or less).  
				\ed 
			\ed 
		\ed 
	\eed   
Some clarifications are in order. First, the appropriate measure of the amount of evidence may depend on the context. In the present case, a good measure is the inverse of the interval length $\ell$; in other contexts, the number of data points might be more suitable. Second, this evaluative standard, like others discussed below, is {\em context-sensitive}, defined with reference to two contextual factors: (i) the question posed and its potential answers, and (ii) the background assumptions in one's inquiry. Third, the `uniform' in `uniform convergence' means that a particular amount of evidence ``works'' across all possible worlds on the table. By the way, `possible worlds on the table' serves as shorthand for `possible worlds compatible with the background assumptions in question.'

Uniform convergence sets an admirably high standard and should be achieved whenever possible. However, it is unachievable in the present context---no inference method for this empirical problem can meet it. The reason is straightforward: an interval of prescribed nonzero length $\ell$ that includes the origin guarantees neither that $\theta$ is $0$ nor that $\theta$ is nonzero, as the true value might be any number within the interval.

Thus, we must lower the bar, at least for this empirical problem. By swapping the two quantifiers `there exists' and `in every' in the standard above, we arrive at a lower standard, traceable to Peirce (1902$a$):
	\oop 
	{\bf Definition (Pointwise Convergence to the Truth).} An inference method $M$ for an empirical problem is said to meet the standard of {\em pointwise convergence to the truth} iff,
		\op 
		\xm in every possible world $w$ on the table, 
			\op
			\xm there exists a prescribed amount of evidence $n$ such that, 
				\op 
				\xm $M$ outputs the answer true at $w$ whenever the amount of evidence reaches or surpasses $n$.  
				\ed 
			\ed 
		\ed 
	\eed  
Informally, this standard requires that, regardless of the true value of $\theta$, the truth must be identifiable {\em at least} in highly favorable situations---with an arbitrarily large but still finite amount of evidence. 

It is routine to show that this lower standard is satisfied by many methods in the present problem, including the method $M^*$, defined as follows: $M^*$ outputs `$\theta = 0$' when the acquired interval includes 0 and `$\theta \neq 0$' otherwise.

We should not settle for this low standard. Instead, we should explore whether the bar can be raised while preserving achievability---a crucial move initiated by Putnam (1965). A recent proposal is particularly relevant as a desideratum for the pursuit of truth, formalizing an ideal admired by Plato toward the end of {\em Meno} (Genin 2018, Lin 2022):
	\oop  
	{\bf Definition (Stability).} An inference method $M$ for an empirical problem is said to be {\em stable} iff, in every possible world on the table, whenever $M$ outputs the true answer, $M$ will never retracts it as the evidence accumulates further. 
	\eed  
Stability, combined with pointwise convergence, establishes an intermediate standard in this hierarchy:
\oop  
\begin{center}
	{\bf Hierarchy for a Non-Stochastic Setting} 
\end{center}
$$\begin{array}{l}
	\quad\quad\quad\quad\quad\quad\quad\quad\quad \vdots
\\
	\textit{Uniform Convergence of the Truth}
\\
	\quad\quad\quad\quad\quad\quad\quad\quad\quad |
\\
	\textit{Pointwise Convergence to the Truth $+$ Stability}
\\
	\quad\quad\quad\quad\quad\quad\quad\quad\quad |
\\
	\textit{Pointwise Convergence to the Truth}
\end{array}$$
\eed  
This intermediate standard is provably achievable in the present problem and is satisfied by the inference method $M^*$ defined earlier. Therefore, in this problem, the bar should be raised to meet this standard. Furthermore, it can be shown that any inference method meeting this standard must conform to a version of Ockham's razor that applies to the short run: whenever the evidence remains compatible with the simpler hypothesis `$\theta = 0$', never accept the more complex hypothesis `$ \theta \neq 0$'---never, ever, including now. For a pictorial proof of a slight variant of this result, see Lin (forthcoming $a$, sec. 3).


This is just one example. Let's step back to consider the bigger picture.


\subsection{The Achievabilist Framework for Convergentism}

The preceding case study suggests a general thesis:
	\oop  
	{\bf Achievabilist Convergentism.} In any empirical problem, an inference method is justified only if it meets the highest achievable mode of convergence to the truth, provided that such a mode exists in the correct hierarchy (pending the specification of the correct hierarchy).
	\eed  
A caveat: this statement is only a first approximation, and complications may arise for various reasons.\footnote{The correct hierarchy might be a partial order (allowing incommensurable standards) or lack a uniquely highest achievable standard. This thesis should be revised accordingly in such cases.} These complications, however, are orthogonal to the present discussion.

This brand of convergentism is quite different from the more traditional varieties, such as Peirce (1902$a$, 1902$b$) and Reichenbach (1938, chap. V), who tend to stick to a specific mode of convergence to the truth. The achievabilist spirit spices up convergentism, allowing context-sensitive flexibility.

Although achievabilist convergentism was articulated in full generality only recently (Lin forthcoming $a$), the idea is not new. It can be traced back to Putnam's (1965) work on using machines to solve mathematical problems (as in computability theory) and was later extended by Kelly (1996) and Schulte (1999) to empirical, non-statistical problems (as in formal learning theory). The same idea also emerged independently in statistics, to which I now turn.




\subsection{Frequentist Hypothesis Testing}\label{sec-freq-testing}

In the empirical problem above, the background assumptions imposes a strong link between the unknown truth and the acquired evidence: every interval as evidence must contain the true value of $\theta$. However, in the context of a statistical problem, the background assumptions are much weaker, rendering the epistemic standards discussed earlier unachievable. The bar must then be lowered again. In simple terms, the ideal of {\em identifying} the truth is replaced by the weaker goal of {\em highly probably identifying} the truth. Let me elaborate.

Here is the new empirical problem:
	\opp 
	{\bf Question (Same as Above).} Is the value of parameter $\theta$ exactly $0$? 
	
	{\bf Background Assumptions (Now Stochastic).} A measurement of $\theta$ produces a point on the real line, being the true (but unknown) value of $\theta$ plus or minus a Gaussian error. Specifically, all data points are generated randomly and independently from a single normal distribution with unit variance, centered on the true value of $\theta$ (commonly referred to as the IID assumption, short for `independent and identically distributed').
	\edd 
An inference method for this problem is still a function: given a finite sequence of data points on the real line as evidence, it outputs one of three options: `$\theta = 0$', `$\theta \neq 0$', or a question mark `\texttt{?}'.

\begin{figure}[ht]
	\centering \includegraphics[width=.7\textwidth]{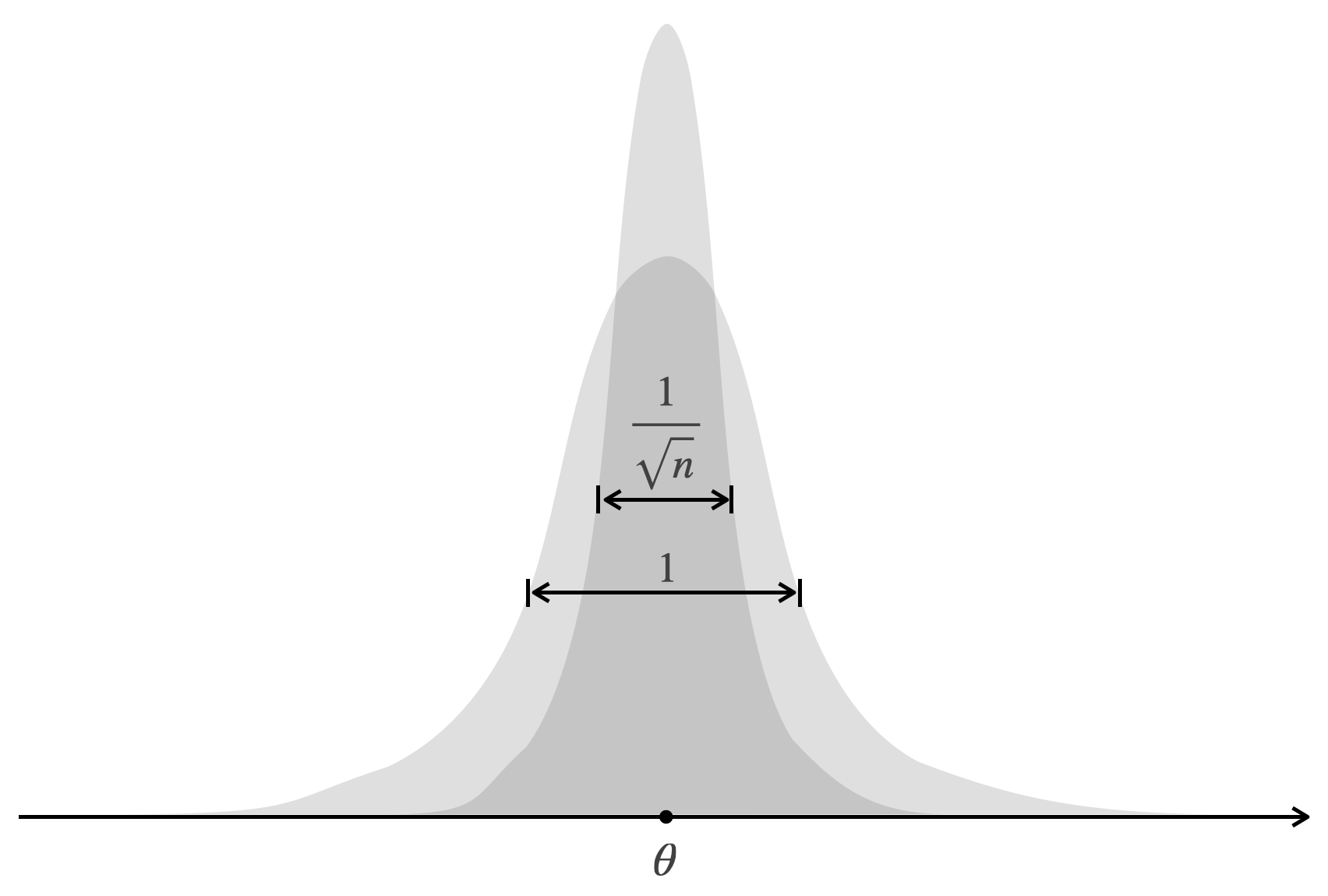}
	\caption{A statistical problem of theory choice: `$\theta = 0$' vs. `$\theta \neq 0$'}
	\label{fig-statistical-testing}
	\end{figure}

Now we need a bit of elementary statistics: In this case, the sample mean is in effect generated from a normal distribution centered on the true value of $\theta$. As the sample size increases from 1 to $n$, this distribution becomes more concentrated around its center, with the spread decreasing by a factor of $1/\sqrt{n}$, as shown in figure \ref{fig-statistical-testing}.	
So, with $n$ measurements, there is a $95\%$ chance that the sample mean lies within $1.96/\sqrt{n}$ of the true value of $\theta$. This motivates an inference procedure for interval estimation: after observing $n$ measurements, construct an interval centered on the sample mean, extending $\pm 1.96/\sqrt{n}$. This procedure has a $95\%$ confidence level, meaning that there is a $95\%$ chance of producing an interval that covers the true value of $\theta$. But how does this help us answer whether $\theta = 0$?

Now, define the inference method $M^\dagger$: let $M^\dagger$ output `$\theta = 0$' whenever the $95\%$ confidence interval contains the origin $0$; otherwise, let it output `$\theta \neq 0$'. This method is a version of Ockham's razor, favoring the simpler hypothesis `$\theta = 0$' unless the data prompt a confidence interval ruling it out. As it turns out, this method is essentially the AIC version of Ockham's razor for model selection, treating `$\theta = 0$' as a model simpler than `$\theta \neq 0$'.\footnote
	{Strictly speaking, when AIC is applied to this problem, it yields the method $M^\dagger$ except for a confidence level of $84.3\%$ rather than $95\%$. See Murtaugh (2014) for details on the tight relationship between AIC and hypothesis testing.}
This method achieves a stochastic mode of convergence to the truth:
	\oop 
	{\bf Definition (Pointwise Convergence to a High Probability).} An inference method $M$ for an empirical problem is said to meet the standard of {\em convergence to a high probability of at least $1- \alpha$ for identifying the truth} iff, in every possible world on the table, the probability that $M$ identifies the truth eventually reaches at least $1-\alpha$ and remains so as the amount of evidence (i.e. sample size $n$) increases indefinitely.
	\eed  
When the high probability threshold $1-\alpha$ is set to $95\%$, this standard is achieved by the inference method $M^\dagger$ as defined above using $95\%$ confidence intervals.

So far, so good, but we should raise the bar if achievability can still be maintained. Instead of fixing a specific high probability threshold, we can impose a high probability of $1-\alpha$ for {\em each} $\alpha > 0$:
	\oop   
	{\bf Definition (Pointwise Convergence to a Probability of $1$.)} An inference method $M$ for an empirical problem is said to meet the standard of {\em pointwise convergence to a probability of $1$ for identifying the truth} iff it meets the previous standard---convergence to a high probability of at least $1-\alpha$---for each $\alpha > 0$.
	\eed   
This higher standard is also provably achievable, but through another version of Ockham's razor: BIC rather than AIC (Nishii 1984). Technical details aside, suffice it to say that, in the present problem, BIC corresponds to an inference method that uses better-designed confidence intervals: as the sample size increases, the interval length still approaches 0, but what sets it apart from AIC is that the confidence level approaches $100\%$ instead of remaining at $95\%$.


Thus, in light of achievabilist convergentism, the BIC version of Ockham's razor holds the upper hand over AIC in the problem of selecting the {\em true} hypothesis (or model), assuming this hierarchy:
\oop  
\begin{center}
	{\bf Hierarchy in Statistical Hypothesis Testing} 
\end{center}
$$\begin{array}{lllc}
	\quad\quad\quad\quad\quad\quad\quad\quad \vdots 
\\
	\textit{Pointwise Convergence to a Probability of 1} 
\\
	\textit{for Identifying the Truth}
\\
	\quad\quad\quad\quad\quad\quad\quad\quad |
\\
	\textit{Pointwise Convergence to a High Probability} 
\\
	\textit{for Identifying the Truth}
\end{array}$$
\eed  


Clarification: While I have presented a distinctively frequentist justification of BIC in the present context, it is worth noting that the `B' in `BIC' stands for `Bayesian'. BIC was originally proposed by Schwarz (1978) in Bayesian statistics, with a justification based on a version of the indifference principle (assigning equal prior credences to the models). I will not assess Schwarz's Bayesian justification here. The point is that, in the present context, BIC, as a learning algorithm independent of its Bayesian interpretation, has a justification grounded in the serious pursuit of truth.

For expository purposes, I have avoided statistical jargon, opting for more descriptive but wordier terms. Here's the translation: The higher standard in the diagram above is referred to as {\em Chernoff-consistency} in frequentist hypothesis testing (Shao 2003, p. 140) and as {\em consistency} in frequentist model selection (Claeskens \& Hjort 2008, ch. 4). Well, consistency in statistics has nothing to do with the consistency of beliefs or peanut butter. The lower standard corresponds to a minor variant of a criterion in frequentist hypothesis testing: {\em a low asymptotic significance level with consistency in power} (Shao 2003, p. 140). For more on the achievabilist aspects of frequentist statistics, see Lin (forthcoming $b$).



I have thus outlined a unified approach to serious pursuit of truth: achievabilist convergentism. Next, I will apply this framework to benefit anti-realists and realists in turn.


\section{Toward an Epistemology for Anti-Realism}\label{sec-instrumentalism}


Let's focus on one of the more traditional anti-realist positions for now: instrumentalism.\footnote{For other varieties of instrumentalism, see the references in Stanford (2016).} Instrumentalists urge pursuit of, not true theories, but useful models; for simplicity I will consider just one aspect of usefulness: predictive power. You will see that there are many valuable results in statistics and machine learning for instrumentalists, and those results can be understood as applications of the epistemological framework I propose: achievabilist convergentism. This point will also be extended to other anti-realist views. 


\subsection{Preliminaries: Parametric Models and Predictive Power}

Imagine an epidemiologist interested in predicting a person's body fat percentage $Y$ based on an easy-to-measure factor $X$, such as BMI (body mass index).\footnote{Defined as body mass divided by the square of height.} An important approach to prediction uses {\em parametric models}. 
A parametric model is a class of functions from $X$ to $Y$ that share a particular form $f(x; \beta_1, \dots, \beta_k)$ with a finite list of adjustable parameters $\beta_1, \dots, \beta_k$. Examples include:
 	$$\begin{array}{ll}
	\text{Polynomial Model of Degree $1$:} & y =  \beta_0 + \beta_1 x \,;
	\\
	\text{Polynomial Model of Degree $2$:} & y =  \beta_0 + \beta_1 x + \beta_2 x^2 \,.
	\end{array}$$
For concreteness, the predictor $X$ and the predicted target $Y$ will be real-valued variables, so that a data set can be visualized as a finite set of points scattered on the two-dimensional $XY$-plane, and a parametric model is a class of curves on the same plane. Generalizations will be discussed when needed. 

Before a model can make predictions, its parameters must be adjusted and fixed to produce a specific curve, usually the one that best fits the available data; the resulting curve $y = f(x)$ on the $XY$-plane is known as the {\em fitted} model. When a new data point $(x_\textrm{new}, y_\textrm{new})$ exists but only $x_\textrm{new}$ is observed, the fitted model makes a prediction of $Y$: $f(x_\textrm{new})$. A measure of the predictive accuracy---or inaccuracy---is often presupposed in a context, such as:
	\begin{eqnarray*}
	\textrm{Absolute Error} &=& \big| y_\textrm{new} - f(x_\textrm{new})\big| \,, \text{ or}
	\\
	\textrm{Squared Error} &=& \big| y_\textrm{new} - f(x_\textrm{new})\big|^2 \,.
	\end{eqnarray*}
While there can be other measures, the key point is that {\em predictive accuracy} is a property that a fitted model has with respect to a data point $(x_\textrm{new}, y_\textrm{new})$.

A fitted model may have high predictive accuracy for one data point and low accuracy for another, and we can define the average of these possible accuracies. More precisely, given a fitted model, its {\em expected predictive accuracy} in a possible world $w$ is defined as a weighted average of its possible predictive accuracies, where the weights are the probabilities true in that same world $w$.

\subsection{Instrumentalist Pursuit of Truth}\label{sec-predictive-model-selection}

A finite sample of data points $D$ on the two-dimensional plane serves two purposes. First, $D$ is used to adjust the parameters of some models to obtain fitted models. Second, the same data set $D$ is used as empirical evidence to help answer this question: 
	\opp 
	{\bf Question in Predictive Model Selection.} Of the fitted models on the table, which one has the highest expected predictive accuracy, or at least comes close to the best in class?
	\edd 
This is a somewhat complicated question, but the idea is straightforward. Start with a simple one: Which fitted model has the highest expected predictive accuracy? It would be ideal to identify the true answer. But if we fail, not all false answers are equally bad. It is preferable to select a model with an expected predictive accuracy close to the best in class. In fact, the same idea is found in the theory of point estimation:
	\opp 
	{\bf Question in Point Estimation.} Which number on the real line is the value of parameter $\theta$, or at least comes close to it?
	\edd 
If the true value is $0.01$, it is still nice to have a false answer that comes close to it, such $0.012$. Not all false answers are equal.

We want a high probability of obtaining an answer close to the true answer to the question posed. This idea, once formalized, is a stochastic mode of convergence to the truth championed by Fisher (1925: sec I.3) in the early days of frequentist statistics:
	\oop  
	{\bf Definition (Pointwise Consistency for Estimation).} An inference method $M$ is said to meet the standard of {\em pointwise consistency for estimation} (or more descriptively, {\em pointwise convergence to a probability of 1 for approximating the truth}) iff, in any possible world on the table, the probability that $M$ outputs an answer $\epsilon$-close to the truth reaches at least $1-\alpha$ and remains so as the sample size increases indefinitely---for any nonzero threshold $\epsilon$ of closeness and for any high probability threshold $1-\alpha$ less than one.
	\eed  
This mode of convergence concerns {\em approximating} the truth, so it sets a lower standard than any modes discussed above, which concern {\em identifying exactly} the truth. 


Now, consider the following hierarchy of modes of convergence to the truth, where {\em asymptotic efficiency} represents an optimal rate of convergence (Shibata 1981), although its exact definition need not detain us here.
\oop  
\begin{center}
	{\bf Hierarchy in Predictive Statistics} 
\end{center}
$$\begin{array}{llc}
	\quad\quad\quad\quad\quad\quad\quad \vdots 
\\	
	\textit{Pointwise Consistency for Estimation}
\\	
	\textit{$+$ Asymptotic Efficiency}
\\
	\quad\quad\quad\quad\quad\quad\quad |
\\
	\textit{Pointwise Consistency for Estimation} 
\\
	\textit{\footnotesize $($i.e. Pointwise Convergence to a Probability of 1 for Approximating the Truth$)$} 
\end{array}$$
\eed  

For simplicity, let the considered models be polynomial. Recall the question posed in this context: {\em Of the fitted models on the table, which one has the highest expected predictive accuracy, or at least comes close to the best in class?} Shibata (1981) proves that, if the background assumption states that {\em none} of the considered models is true (i.e., the true curve is not a polynomial), it is possible to achieve the higher mode of convergence in the above diagram, and it is achieved by the AIC version of Ockham's razor, rather than BIC.

The preceding result is highly context-sensitive. Suppose that the context switches to one in which the same question is posed but the background assumption states, instead, that {\em some} of the considered models is true. Then the same mode of convergence is still achievable, but this time it is achieved by BIC rather than AIC (Shao 1997)---the recommendation is reversed.

Here is another example: What if we stick to the same question that interests instrumentalists, but the background assumptions are now weaker, saying nothing about whether one of the considered models is true? This has motivated a research program to find a new version of Ockham's razor, sharper than AIC but blunter than BIC; see the survey by Ding et al. (2018). 

To dramatize context-sensitivity, recall an earlier example: What if we switch to a question that generally incurs disapproval from instrumentalists, such as which polynomial models on the table are true, assuming that some of them is true? In this context, the recommended method of model selection is BIC, which achieves a higher mode of convergence than AIC does (Nishii 1984) as seen above (section \ref{sec-freq-testing}). So, whether the posed question makes instrumentalists happy or unhappy, the underlying epistemology can be the same: achievabilist convergentism. 


\subsection{Prospects for Anti-Realist Epistemology}

The achievabilist framework is general enough to accommodate other anti-realists. 

While we have seen that Shibata (1981) and Shao (1997) directly tackled the model selection question, the more traditional approach, due to Akaike (1973), takes a slight detour. He first addressed an estimation question before applying it the model selection question:
	\opp 
	{\bf Akaike's First Question: Estimation.} What is the predictive power of this or that fitted model?
	\edd 
	\opp 
	{\bf Akaike's Second Question: Selection.} Of the fitted models on the table, which one has the highest predictive power, or at least comes close to the best in class?	 
	\edd 
To address the first, estimation question, Akaike (1973) developed an estimator, known as the {\em AIC score}, which uses the available data to estimate the expected predictive accuracy of a fitted model by this quantity: the degree of fit with the available data minus (i.e., penalized by) the degree of complexity (measured by the number of parameters). Akaike (1973) proved that, under quite general assumptions, the AIC score as an estimator has an important property known as {\em asymptotic unbiasedness}, which is only a necessary condition for meeting the standard of pointwise consistency.\footnote
	{Relative to a sample size held fixed, the {\em bias} of an estimator is the expected value of its estimate minus the true value of the estimated quantity. An estimator is called {\em asymptotically unbiased} if its bias converges to $0$ as the sample size increases indefinitely. In order for such an estimator to meet the standard of pointwise consistency for estimation, one more condition is needed: the ``spread'' of the possible estimates generated by this estimator, known as the {\em variance} of this estimator, must also converge to $0$ as the sample size increases indefinitely.} 
Akaike's followers have since pursued pointwise consistency for the estimation problem and, in recent years, pushed the boundary of achievability to attain a high rate of convergence (Austern et al. 2020; Bayle et al. 2020; Wager 2020; Li 2023)---their endeavors are clearly achievabilist at least in spirit, if not in name. For a survey of those results aimed at a philosophical audience, see Lin (2024$a$, sec. 5). 

Many machine learning researchers are interested in the same questions about predictive power, too, except that they often need to incorporate binary variables. Consider, for example, the task of predicting whether a $2048 \times 2048$ image depicts a cat based on its more than 4 million pixels. Here, the value of the predictor variable $X$ can be any such image, that is, a vector in a more than 4 million dimensional space. The predicted variable $Y$ is binary: $Y = 1$ means that it is a cat, $Y = 0$ means that it is not. In image classification, the most popular models are {\em neural networks}; they are still parametric, and their parameters are called {\em weights}, representing the strengths of signal transmission between artificial neurons. The corresponding estimation problem was pioneered by machine learning theorists Vapnik \& Chervonenkis (1971) and Vapnik (2000, chap. 4), who adopt interval estimation in contrast to Akaike's use of point estimation. The corresponding model selection problem was addressed by machine learning theorists Lugosi \& Zeger (1995, 1996) and Vapnik (2000, chap. 4). The theme is the same: taking pointwise convergence as the minimum qualification, those researchers explored higher and higher achievable modes of convergence to the truth---the true answer to the question posed. For a textbook presentation of this field, known as {\em statistical learning theory}, see Shalev-Shwartz \& Ben-David (2014, part I). 

So much for instrumentalists---but these points carry over to other varieties of anti-realism. Consider constructive empiricists (van Fraassen 1980), who urge the pursuit of an {\em empirically adequate} theory, a theory that only has true consequences about observations. As such, constructive empiricists should be interested in true answers to questions of this form:
	\opp 
	{\bf Constructive Empiricists' Question.} Is it the case that the considered theory's consequences about observations are all true?
	\edd 
For example, consider ``all ravens are black'' as a theory. Then the question becomes this: {\em Is every raven observed, past or future, black?} Or consider a more interesting theory, and enumerate all of its consequences about observations: $O_1, O_2, \ldots$; then ask: {\em Are those (countably many) observational consequences, $O_1, O_2, \ldots$, true?} Such empirical questions are actually the core objects of study in formal learning theory (Kelly 1996, Schulte 1999, Kelly 2011, and Kelly, Genin, \& Lin 2016). The theme is the same: starting with pointwise convergence, and exploring higher and higher achievable modes of convergence to the truth---the true answer to the question posed---even though different anti-realists might pose different questions.

Thus, the achievabilist framework for convergentism allows anti-realists to leverage rich resources available in statistics, machine learning, and formal epistemology. Next, I want to deliver the same promise to realists.



\section{Toward an Epistemology for Realism}\label{sec-realism}

We have seen that anti-realists can be united under the framework of achievabilist convergentism together with this addendum: whatever the correct hierarchy is for modes of convergence as evaluative standards, the minimum qualification is {\em pointwise} convergence to the truth (stochastic or not). Recall that, here, `pointwise' means `at each point.' So, the minimum qualification for anti-realists is convergence to the truth at {\em each} possible world on the table. Unfortunately, even this relatively low standard is unachievable in the empirical problems that matter to realists---due to a particularly severe type of underdetermination of theory by data, as we will see shortly. 

A natural response from realists---in fact, a distinctively achievabilist one---is to lower the bar further and explore the possibility of convergence at {\em almost all} (rather than all) possible worlds on the table. I will flesh out this idea, drawing on recent advances in machine learning and formal epistemology, and applying it to an iconic example: Einstein's and Perrin's case for atomism. 




\subsection{Preliminaries: Einstein and Perrin on Atomism}

When pollen particles are suspended in a fluid (e.g., water), they exhibit random, zig-zag motion known as Brownian motion. Einstein (1905, 1907) showed that if atomic theory is true (together with certain auxiliary hypotheses), this motion results from pollen being bombarded by molecules in the fluid. This explanation involves the Avogadro number $N_{\!A}$, which measures the granularity of matter and denotes, say, the number of molecules in 32 grams of oxygen gas under atomic theory. A smaller $N_{\!A}$ implies coarser fluid division, heavier molecules, stronger molecular impacts, and more drastic pollen motion. Einstein formalized this idea, deriving a negative correlation between $N_{\!A}$ and the displacement of pollen motion: 
\begin{eqnarray*} 
\textbf{(Einstein's Equation)} \quad \textit{Mean Squared Displacement} &=& \frac{c}{N_{\!A}} \cdot \textit{Time} \,, 
\end{eqnarray*} 
where the mean in {\em Mean Squared Displacement} is over pollen particles, and $c$ is an algebraic combination of observables. Since all quantities except $N_{\!A}$ are observable, this equation allows $N_{\!A}$ to be estimated under atomic theory. 

Perrin used atomic theory to explain a distinct phenomenon: like barometric pressure, the density of gamboge particles in a vertical fluid column decreases exponentially with height. More specifically, Perrin derived:
\begin{eqnarray*} 
\textbf{(Perrin's Equation)} \quad \textit{Vertical Density} &=& e^{-c' \cdot N_{\!A} \cdot \textit{Height} } \,, \end{eqnarray*} 
where $c'$ is another combination of observables. Like Einstein's equation, this allows $N_{\!A}$ to be estimated. Atomic theory implies the equality between the values of $N_{\!A}$ from Einstein's and Perrin's equations. Testing this equality provides a way to falsify atomic theory.

There are other ways to falsify atomic theory, such as by testing the linearity of Einstein's equation and the exponentiality of Perrin's. But for simplicity, let's assume God has told us their linearity and exponentiality. Then our focus is the same as Perrin's: testing the equality between the two copies of the Avogadro number in the two equations.

Perrin, like any realist, faced a particularly severe type of underdetermination by data. Let me visualize it with a model. 

\subsection{A Pasta Model of Underdetermination}

\begin{figure}[ht]
	\centering \includegraphics[width=.7\textwidth]{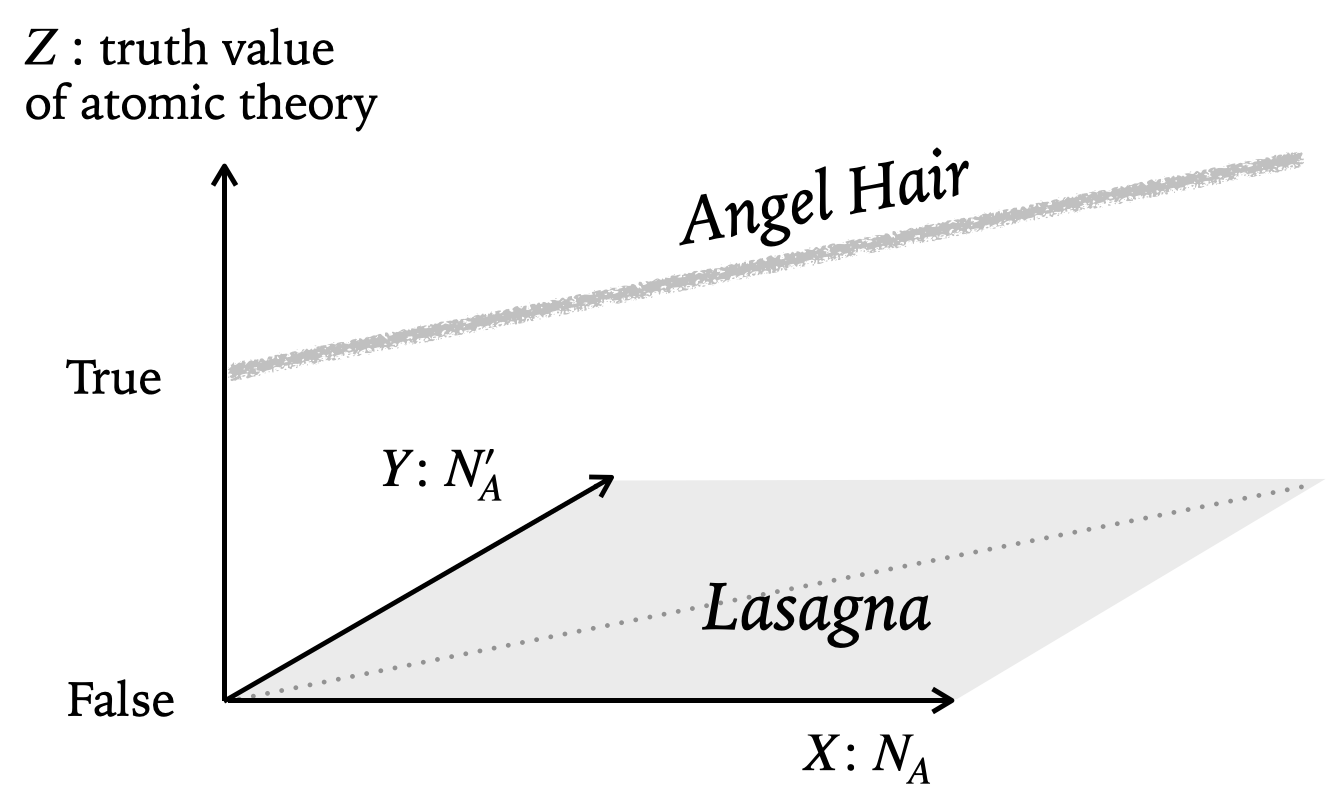}
	\caption{A pasta model of Perrin's case}
	\label{fig-pasta}
	\end{figure}

Take a look at figure \ref{fig-pasta}, which features a strand of angel hair pasta floating above a sheet of lasagna.
The $X$-axis represents the Brownian motion of pollen particles in a fluid, simplified to a single free parameter, $N_{\!A}$, which appears in Einstein's equation: 
	\begin{eqnarray*}
	\textit{Mean Squared Displacement} 
		&=& \frac{c}{N_{\!A}} \cdot \textit{Time} \,.
	\end{eqnarray*}
The $Y$-axis represents the vertical density distribution of gamboge particles in a fluid, simplified to another free parameter, $N'_{\!A}$, in Perrin's equation: 
	\begin{eqnarray*}
	\textit{Vertical Density}
		&=& e^{-c' \cdot N'_{\!A} \cdot\textit{Height} } \,.
	\end{eqnarray*}
It's important to keep $N_{\!A}$ and $N'_{\!A}$ distinct, as the equality between those two is a point of contention in the present context. Perrin conducted experiments to test whether their values are the same. Indeed, their values might need to be set different in order for Einstein's and Perrin's equations to fit data well, serving as good phenomenological models. 

Now consider the $Z$-axis, which indicates whether atomic theory is true. $Z = 1$ means that the theory is true; $0$, false. So, atomic theory is false at the possible worlds on the $XY$-plane ($Z = 0$), forming a sheet of lasagna pasta, where the two Avogadro parameters can have the same or distinct values. In contrast, atomic theory implies that $N_{\!A} = N'_{\!A}$; it is true at the possible worlds with $Z = 1$ along the diagonal, forming a strand of angel hair pasta above the lasagna sheet. These represent all the possible worlds on the table---the worlds compatible with the background assumptions.

I encourage you to adapt this picture to any culinary culture of your choice, such as switching from Italian to Vietnamese cuisine, using a strand of rice vermicelli and a sheet of rice paper instead.

\begin{figure}[ht]
	\centering \includegraphics[width=.7\textwidth]{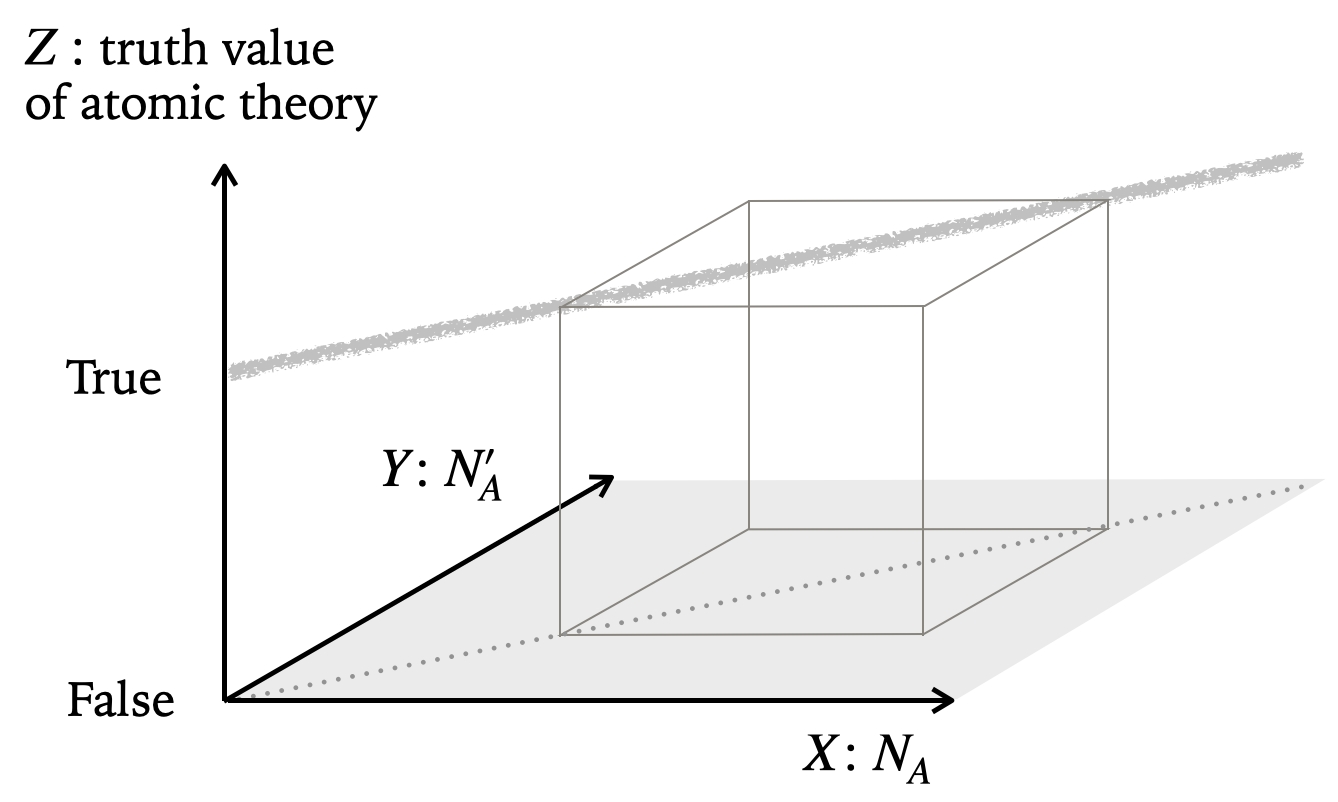}
	\caption{An example of evidence in Perrin's case}
	\label{fig-prism}
	\end{figure}

A body of empirical evidence consists of measurements of $N_{\!A}$ and $N'_{\!A}$, simplified as intervals on the $X$- and $Y$-axes. Such evidence, assumed to always contain the true values (as in formal learning theory), rules out some possible values of $N_{\!A}$ and $N'_{\!A}$ but doesn't logically distinguish between the possibilities along the $Z$-axis. Thus, a body of evidence can be depicted as a rectangular prism, ruling out the worlds outside it, as depicted in figure \ref{fig-prism}.

\begin{figure}[ht]
	\centering \includegraphics[width=.7\textwidth]{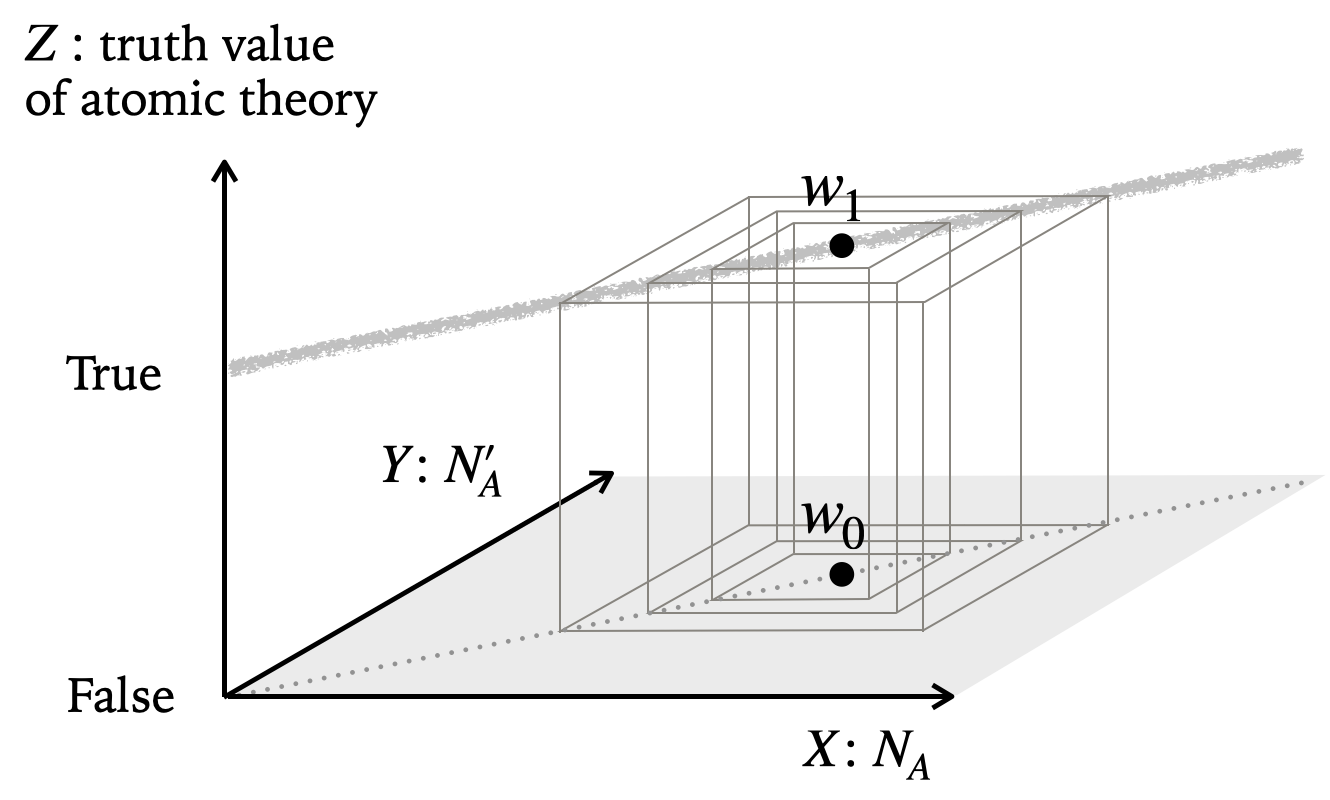}
	\caption{Underdetermination in Perrin's case}
	\label{fig-prism2}
	\end{figure}

Now, realists pose this question: 
	\opp 
	{\bf The Realist Question in Perrin's Case.}
	Is atomic theory true? In other words, is the actual world on the strand of angel hair or on the lasagna sheet?
	\edd 
Severe underdetermination then arises. The strand of angel hair has a projection onto the $XY$-plane---an {\em empirically equivalent} counterpart within the lasagna sheet, where the two Avogadro parameters {\em happen} to be identical. For example, as depicted in figure \ref{fig-prism2}, the possible world $w_1$ on the angel hair strand has an empirically equivalent counterpart $w_0$ on the lasagna sheet.
Any data sequence produced in one of the two worlds, such as the sequence of nested prisms shown in that figure, could have been produced in the other. Here is the crux: an inference method behaves the same when fed a particular data sequence regardless of whether this data sequence is produced in world $w_0$ or $w_1$. So, to converge to the true answer in one of those two worlds, $w_0$ or $w_1$, is to convergence to the same answer---a falsehood---in the other world. Hence, in this problem context, it is impossible to converge to the truth everywhere in the space of the possible worlds on the table---there is no way to meet the standard that the anti-realists in statistics and machine learning would take as the minimum qualification.\footnote{This impossibility also extends to the setting where data are generated stochastically. In the stochastic setting, worlds $w_0$ and $w_1$ share the same probability distribution over the set of possible data sequences, making it impossible for any inference method to achieve stochastic convergence to the truth in both worlds.} 

Thus, if realists wish to stay in the framework of achievabilist convergentism, they must lower the minimum qualification and explore the possibility of {\em almost} everywhere convergence. This has been done only recently in machine learning and formal epistemology, albeit in the context of learning causal relations from statistical data, due to Lin (2019) and Lin \& Zhang (2020).\footnote{Their work is build upon Spirtes, Glymour, \& Scheines (2001) and Meek (1995) on how the causal faithfulness assumption holds almost everywhere.} Those authors propose to leverage geometers' definitions of ``almost everywhere.''

To that end, I need to provide a brief tutorial on a branch of geometry.

\subsection{Topology Without Tears}

In the pasta model, some geometric information matters, but not all details do. For example, the exact Euclidean distance between points is irrelevant, as there may be no philosophically meaningful metric between possible worlds. This model can be stretched along any dimension. What matters is whether a set of worlds {\em comes arbitrarily close} to a world---like the open interval $(0, 1)$ comes arbitrarily close to $1$ but not $1.001$. 

Topology studies the mathematical properties that can be defined by the concept of arbitrary closeness (or equivalently, the concept of open sets).\footnote
	{
	While topology is typically presented with ``open sets'' as the basic notion, I take ``coming arbitrarily close'' as basic in this context for elegance. Those two notions are inter-definable in topology (Arkhangel'skii \& Fedorchuk 1990, secs. 1.1-1.4). A set $S$ is said to {\em come arbitrarily close} to a point $p$ iff every open set containing $p$ overlaps with $S$. A set $S$ is called {\em open} iff the complement of $S$ comes arbitrarily close to no member of $S$. 
	}
For example, a function from one space of points to another is called {\em continuous} iff it preserves the relation of arbitrary closeness---stretching a shape without tearing, though it allows gluing. 


The pasta model misses metric information, but it faithfully represents the topological structure---the relation of arbitrary closeness. Consider the possible world $w_1$ where atoms exist and the Avogadro number is, say, $6.02 \cdot 10^{23}$, as in figure \ref{fig-prism2}. The geometry of the pasta model suggests that $w_1$ can be approximated by a sequence of points on the angel hair strand, but not by the points on the lasagna sheet. This geometry captures the intuitive relation of arbitrary closeness among possible worlds: to approximate $w_1$, where atoms exist and the Avogadro number is $6.02 \cdot 10^{23}$, we must use worlds where atoms still exist, adjusting the Avogadro number. Atomless worlds are very different from worlds with atoms; the former do not come arbitrarily close to the latter.

Several standard geometric definitions of ``almost everywhere'' exist, some more stringent than others. For epistemological purposes, we should try out the most stringent definitions first, unless they all are unachievable. The most stringent definitions all adhere to the following two principles:\footnote
	{Here are the two most stringent definitions of ``almost everywhere'', neither of which is more stringent than the other:
	\op 
	\im[(1)] The first is purely topological: being large enough to include a dense open set in a topological space. 
	\im[(2)] The second is measure-theoretic: being large enough to exclude only a set of {\em Lebesgue measure zero} in a finite-dimensional Euclidean space, or more generally, to exclude only a set of {\em Haar measure zero} in a locally compact topological group. 
	\ed 
	For each of the definitions mentioned above, denseness is a necessary condition. Yet denseness is not a necessary condition for a less stringent definition, such as that of {\em comeager} sets, which only requires inclusion of a countable intersection of dense open sets. See Lin (2019, 2022) for an accessible presentation of the purely topological approach written for philosophers. The measure-theoretic approach is still somewhat topological, as it presupposes a topology. See Oxtoby (1980) for a now-classic comparison of the purely topological and the Lebesgue-measure approaches. See Diestel et al. (2014) for a comparison of all the aforementioned approaches, including the Haar-measure approach. 
	}
	\oop   
	{\bf The Denseness Principle.} A region covers almost everywhere in a space {\em only if} the former is dense in the sense that it comes arbitrarily close to every point in the latter.
	\eed   
	\oop   
	{\bf The Lower Dimension Principle.} A region covers almost everywhere in a space {\em if} the former misses only a point in a line, or a line in a plane, or a lower dimensional region in a higher dimensional one.
	\eed   
Although it is not obvious, the concept of dimensions is also topological, in the sense that the dimension of a space is invariant under any one-to-one continuous mapping to and from another space---continuous in both directions, ensuring that no tearing or gluing occurs.\footnote
	{This is Brouwer's {\em invariance of domain} theorem; see Fedorchuk (1990, pp. 95-97).} 

End of the tutorial on topology. Let's return to philosophy. 

\subsection{Striving for ``Almost Everywhere''}

It is time to tackle Perrin's empirical problem, which poses the question whether atomic theory is true. For clarity, we need some definitions:
	\oop   
	{\bf Definition (Convergence to the Truth, Domain of Convergence).} An inference method $M$ for an empirical problem {\em converges to the truth} at a possible world $w$ on the table iff there exists an amount of evidence $n$ such that, whenever the given evidence in world $w$ reaches or surpasses the amount $n$, method $M$ outputs the answer true in world $w$. The set of such worlds on the table is called $M$'s {\em domain of convergence}.
	\eed   
This allows for a streamlined definition of various modes of convergence:
	\oop 
	{\bf Definition (Everywhere, Almost Everywhere, Maximal, Stability).} 
	\op 
	\im {\em Everywhere convergence}, which is the same as pointwise convergence, means that the domain of convergence covers the entire space of the possible worlds on the table. 
	\im {\em Almost everywhere convergence} means that the domain of convergence covers almost everywhere in the space of the possible worlds on the table. 
	\im A domain of convergence is called {\em maximal} iff no inference method's domain of convergence properly includes it.
	\im {\em Stability}, let's recall, means that, in every possible world on the table, the true answer, once inferred, is never retracted when more evidence is acquired.
	\ed    
	\eed  
Now we can construct the following hierarchy of modes of convergence:
\oop  
\begin{center}
	{\bf An Extended Hierarchy} 
\end{center}
$$\begin{array}{l}
	\quad\quad\quad\quad\quad\quad \vdots 
\\
	\textit{Everywhere Convergence} 
\\
	\quad\quad\quad\quad\quad\quad |
\\
	\textit{Almost Everywhere Convergence} 
\\
	\textit{$+$ Maximal Domain $+$ Stability}
\\
	\quad\quad\quad\quad\quad\quad \vdots 
\end{array}$$ 
\eed  

Let's examine some inference methods and see what they achieve. 

First, consider the inference methods following this methodological principle, called {\em Ockham's realist razor}: 
	\op 
	\im {\sc (Long-Run Realist)} Converge to the truth at every simple world, where atomic theory is true, thereby sacrificing convergence at their empirically equivalent counterparts.
	\im {\sc (Short-Run Ockham)} Whenever the evidence is still compatible with the simple hypothesis (atomic theory), never infer the complex hypothesis (the negation of atomic theory)---never, ever, including now.
	\im {\sc (Deduction)} Whenever the evidence is incompatible with the simple hypothesis, deductively infer the complex hypothesis (under the background assumptions).
	\ed
Then we can prove the following result, presented as a score sheet:
\oop  
\begin{center}
	{\bf Theorem: Score Sheet in Perrin's Case} 
\end{center}
$$\begin{array}{llll}
	\quad\quad\quad \vdots 
\\
	\textit{Everywhere Convergence} 
	&&& \text{This is unachievable.}
\\
	\quad\quad\quad |
\\
	\textit{Almost Everywhere Convergence} 
	&&& \text{This is achieved by, and only by,}
\\
	\textit{$+$ Maximal Domain $+$ Stability}
	&&& \text{{\bf Ockham's realist razor}.}
\\
	\quad\quad\quad \vdots 
	&&& \text{All the other methods, including}
\\
	&&& \text{the {\bf anti-realist method},}
\\
	&&& \text{fail to achieve that much.}
\end{array}$$ 
\eed  
\noindent I will provide the most interesting part of the proof here, leaving the remaining part to the appendix.

\begin{figure}[ht]
	\centering \includegraphics[width=.7\textwidth]{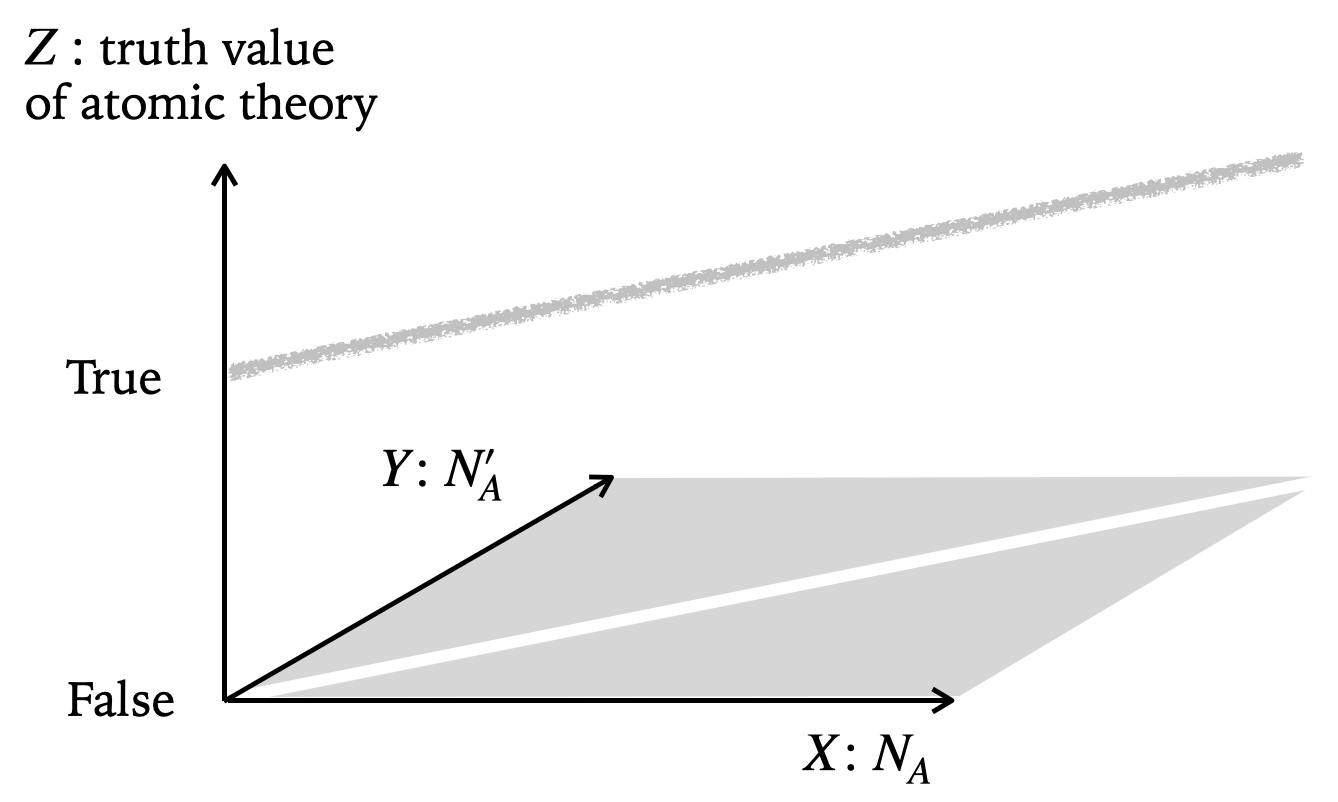}
	\caption{The gray region is where Ockham's realist razor converges to the truth}
	\label{fig-realist}
	\end{figure}

It is routine to show that any method following Ockham's realist razor converges to the truth in exactly the gray region depicted in figure \ref{fig-realist}. 
That is, it converges to the truth at every simple world (where atomic theory is true), sacrificing convergence only at the complex worlds empirically equivalent to a simple world (on the diagonal in the lasagna sheet), while converging to the truth at all other complex worlds. So, the domain of convergence misses only a line in a plane; then, by the Lower Dimension Principle, it covers almost everywhere in the entire space. It is easy to see that this domain of convergence is maximal as it cannot be extended further (due to the sort of severe underdetermination in question). It is also routine to verify the stability property. So, any inference method following Ockham's realist razor achieves the second standard displayed in the score sheet, which is then the highest achievable, as we have seen that the first one is unachievable. 

We can further show that any other methods violate the second standard in the score sheet---failing either almost everywhere convergence, maximal domain, or stability. The full proof of this part is in the appendix. But for here, let me walk you through a case of particular interest.

\begin{figure}[ht]
	\centering \includegraphics[width=.7\textwidth]{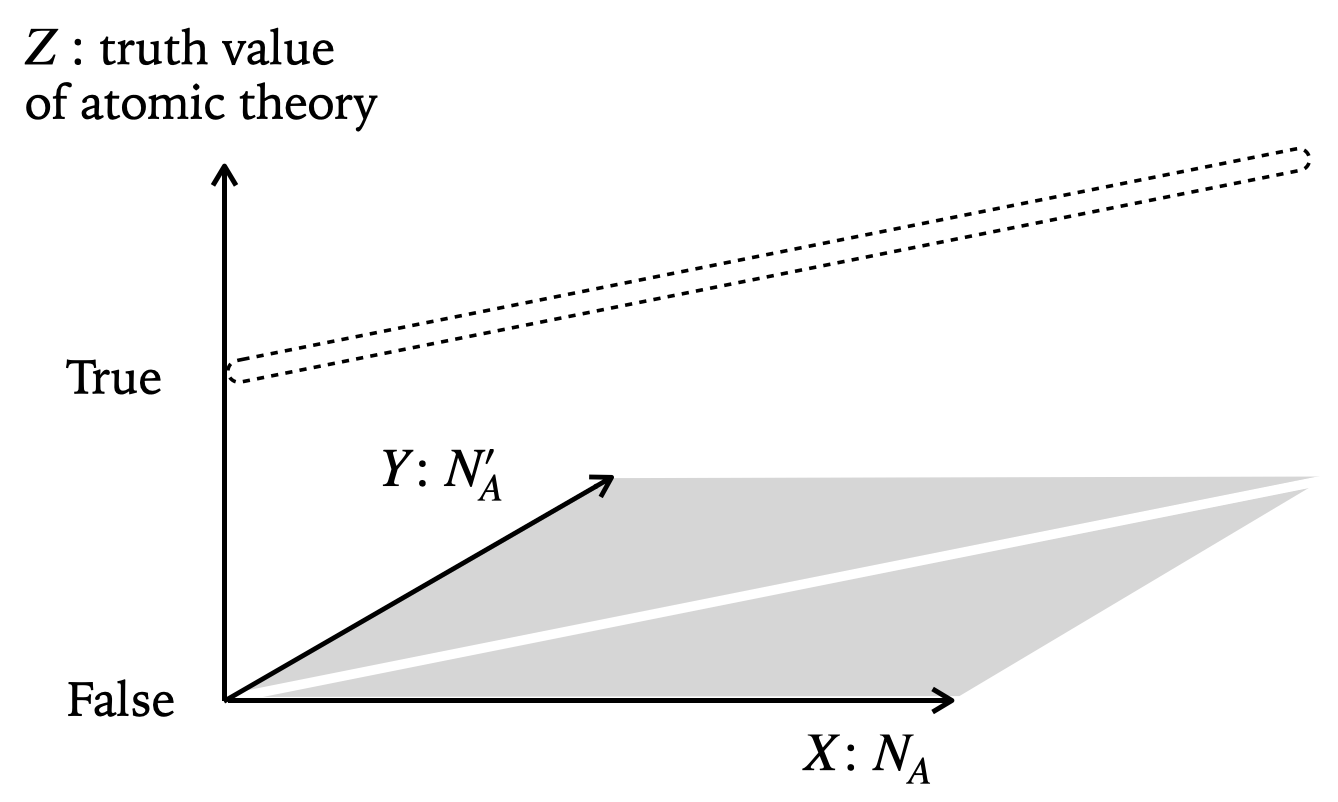}
	\caption{The gray region is where the anti-realist method converges to the truth}
	\label{fig-anti-realist}
	\end{figure}

The {\em anti-realist method} is defined as follows:
	\op 
	\im Remain agnostic about the truth or falsity of atomic theory whenever the evidence is compatible with both. 
	\im Whenever the evidence is incompatible with atomic theory, infer its negation.
	\ed
It follows that this method fails to converge to the truth on the entire strand of angel hair as well as its counterpart on the lasagna sheet---depicted as the gray region in figure \ref{fig-anti-realist}.
So, the anti-realist's domain of convergence fails to be maximal, for there exist methods with a strictly more inclusive domain of convergence, covering the angel hair strand. Moreover, the anti-realist's domain of convergence fails to come arbitrarily close to some possible worlds, such as those on the angel hair strand; then, by the Denseness Principle (``almost everywhere'' implying ``coming arbitrarily close to every point''), the anti-realist's method fails the criterion of almost everywhere convergence. Upshot: the anti-realist's method does not meet the second standard in the score sheet, failing two of the three criteria therein.


This concludes the most interesting part of the proof for the theorem; the remainder is provided in the appendix.

Thus, in Perrin's context, the highest achievable standard is almost everywhere convergence with maximal domain and stability, and it is achieved by, and only by, the methods following Ockham's realist razor. This explains how realist-friendly inference methods connect to truth in Perrin's case---they are the only methods achieving the highest achievable mode of convergence to the truth. Moreover, this explanation is non-circular: the conclusion is derived without applying Ockham's razor or IBE, and the key concepts in use are independently motivated, such as ``almost everywhere'' from geometry, and stability from Plato's {\em Meno}. 

To be sure, this result would not impress anti-realists---and I have no intention to do that. The point is, rather, to address those who are still undecided about the realism debate by attempting a non-circular account of how realist-friendly inference connects to truth.

The above is only an initial step toward a realist epistemology without circularity, though. It comes with some restrictions: it only concerns Perrin's case, employs a quite simplistic modeling of Perrin's epistemic situation, and restricts itself to a non-stochastic setting for evidence generation. To obtain a robust realist epistemology, those restrictions need to be relaxed, possibly in a piecemeal way, and I remain optimistic: the concept of stability and its crucial role for justifying Ockham's razor have been extended by Genin (2018) to a stochastic setting; Lin \& Zhang (2020) and Lin (2022) have obtained some general theorems to leverage the concept of almost everywhere convergence, in both stochastic and non-stochastic settings. 

I recognize that some realists might still wish to have a sweeping account of how their beloved inference connects to truth---a one-stop global solution covering all cases in a single argument, like Putnam's and Psillos's IBE-based arguments that IBE is truth-conducive. The cost they incur, as noted above, is circularity. It is the intention to avoid circularity that I turn away from the one-stop solution and go {\em local} instead: examining one problem context at a time (such as Perrin's case) and paying attention to the details therein in order to do something positive for realism. In this regard, I join the localist approach to scientific realism (Magnus \& Callender 2004, Fitzpatrick 2013, Saatsi 2017). I hasten to add that the localist approach fits nicely with the epistemology I propose, achievabilist convergentism, which is a general framework that takes extra care of context-sensitivity. The operative standard for assessing inference methods is context-sensitive, and set to be the highest achievable mode of convergence in {\em each} problem context. 

Enough with nice words to realists. It is time to step back.
 



\section{In Search of Deeper Disagreements}\label{sec-deepest}

There are at least three aspects to the realism debate.\footnote
	{See Chakravartty (2017, sec. 1.2) for a slightly different categorization of three aspects: metaphysical, semantic, and epistemological. I do not discuss the metaphysical aspect here but instead consider the teleological aspect. Strictly speaking, there should be four aspects in total.} 
The first is {\em semantic}, concerning whether realists argue that scientific theories should be interpreted literally.
	The second aspect is {\em teleological}: realists are often understood to maintain that science aims at truth, while anti-realists are often portrayed as attributing a different aim to science.
		The third aspect, which has been the focus of this paper, is {\em epistemological}: realists contend that, given appropriate empirical evidence, we are justified in believing in the (mind-independent) existence of entities like atoms and possibly other unobservables, while anti-realists argue that no empirical evidence can justify such beliefs.

Those aspects of the debate---semantical, teleological, and epistemological---are not equally important. The epistemological one strikes me as the most fundamental. Moreover, a deep epistemological disagreement emerges only through the lens of the common ground I have developed for both realists and anti-realists. Let me explain.

\subsection{The Primacy of the Epistemological}


The semantic aspect of the debate seems to me the least fundamental of the three. Indeed, some anti-realists (van Fraassen 1980) join realists in thinking that scientific theories should be interpreted literally, while other anti-realists disagree (Rowbottom 2011). This suggests that the semantic debate, though highly relevant, is not central.

The semantic debate is non-fundamental for a more important reason. Instead of arguing that a theory {\em should} be interpreted in one way rather than the other way, let's be liberal: a theory {\em may} be interpreted in {\em either} way---any theory, under any clear interpretation, can be a choice on the table. So there can even be a choice between two alternative interpretations of the same theory; it is a choice between two interpreted theories. Just be explicit about any considered interpretations, and make sure that any theory choice is a choice among clearly interpreted theories. So, the semantic debate on the proper interpretation seems to be not as central as the question of theory choice---choosing from among clearly interpreted theories with respect to a certain goal. But wait: what goal is the goal to pursue? 

This naturally leads to a more fundamental, teleological aspect of the debate. Realists hold that the aim of science is to find truth, making the goal of theory choice to select a true or approximately true one. In contrast, anti-realists maintain that the aim of science is to choose a theory or model that has certain properties other than truth, such as empirical adequacy (van Fraassen 1980, 1994), or predictive power (Sober 2002).

Underlying the teleological issue, there seems to me something even more fundamental. Here is the idea. Once the goal is to find a theory or model with contingent property $X$---{\em whatever $X$ may be}---a question arises: 
	\opp 
	{\bf Empirical Question.} Which theory or model on the table has contingent property $X$, if any? 
	\edd 
This, in turn, raises a meta-question:
	\opp 
	{\bf Epistemological Question}: What empirical evidence would strongly support this or that potential answer to the empirical question posed above? 
	\edd
This epistemological question is important for the choice of a goal. Anti-realists generally worry that no empirical evidence is strong enough to justify a belief in the existence of atoms. If so, it makes no sense to pursue the question of whether atoms exist, as one of the potential answers---`yes'---could never be strongly supported by evidence. This epistemological qualm seems to be the deeper reason that anti-realists have for rejecting the teleological claim in realism. This line of thought suggests the following principle:
	\oop  
	{\bf Epistemological Criterion of Sensible Goals.} In any problem context, the goal to pursue a theory with contingent property $X$ is sensible only if the following condition holds: for every theory $T$ on the table, there exists a possible body of empirical evidence that strongly supports, and thus justifies a belief in, the proposition that theory $T$ has property $X$.
	\eed  
So, whether a goal makes sense depends in part---but crucially---on the corresponding epistemological meta-issue. Underlying the teleological issue, the deeper issue is the epistemological issue of evidential support and justified belief. 



\subsection{One Step Deeper}

I suspect that the disagreement on the strength of evidential support is not even a deepest issue. Evidence, if anything, seems to be an indicator of truth by its very nature. So, a deeper disagreement seems to lie in the issue of how evidential support connects to truth---the issue of {\em truth connection}. To clarify this issue (rather than resolving it), I propose a two-step connection, from evidential support to a waypoint before reaching truth:
	\oop  
	{\bf Bridge I (From Support to Inference).} In any empirical problem, $E$ strongly supports $H$---to an extent that justifies a belief in $H$ given evidence $E$---iff there exists a justified inference method that outputs $H$ given $E$. 
	\eed  
	\oop  
	{\bf Bridge II (From Inference to Truth; Achievabilist Convergentism).} In any empirical problem, an inference method is justified only if it achieves the highest achievable mode of convergence to the truth, provided that such a mode exists in the correct hierarchy ({\em pending the specification of the correct hierarchy}). 
	\eed  
The second of these two bridge principles is exactly achievabilist convergentism, and the demon is in the parentheses. As we have seen in previous sections, anti-realists would hold that the correct hierarchy only extends down to the standard of {\em everywhere} convergence to the truth (stochastic or not), being the minimum qualification---no lower standard is worth striving for. Realists, on the other hand, would argue that the correct hierarchy extends further below, down to the standard of {\em almost everywhere} convergence to the truth. In other words, the crux of the matter is this: {\em What is the minimum qualification for the link between justified inference and truth, everywhere convergence or almost everywhere?} This seems to me a deepest divide between the two parties. If there cannot be a good argument that favors one particular cut-off instead of the other, the realism debate is irreconcilable.

The disagreement just presented is distinctively epistemic: it pertains to truth, or to modes of convergence to the truth, and to almost nothing else. In particular, this disagreement was just formulated with reference to the value of scientific explanation, which some take to be another source of the irreconcilability (Chakravartty 2017, ch. 7; Forbes 2017). As a value that realists generally champion and anti-realists deny, the value of explanations appears to be a plausible candidate for one of the roots of the irreconcilability, but it is not clearly epistemic. Indeed, many anti-realists argue that the value of explanations, if any, is pragmatic rather than epistemic (van Fraassen 1980). In contrast, achievabilist convergentism as a common ground helps us isolate a distinctively epistemic root of the irreconcilability, whether or not there are multiple roots. 

\section{Closing}\label{sec-irreconcilability}

Van Fraassen (1989, ch. 5) offers another notable explanation of the irreconcilability: realists and anti-realists can both be rational, as rationality is only a matter of internal coherence.\footnote{To clarify, van Fraassen explanation has been incorporated by Chakravartty (2017, ch. 7) and Forbes (2017) into the value-based explanation, regarding especially the value of explanations as discussed above.} The explanation I sought is different, shifting the focus from rationality to justification, and from internal coherence to truth finding. My proposal, to recap, is that realists and anti-realists can (and should) both love truth, agreeing that justified inference must have {\em some} connection to truth---but {\em what} connection? How strong the connection to truth must there be in the very least? Or put more formally, which mode of convergence to the truth is the minimum qualification: everywhere convergence or almost everywhere convergence? If there is no good argument for a particular cut-off, the irreconcilability follows.


Van Fraassen's explanation of the irreconcilability and mine embodies two opposing approaches to epistemology of science. His explanation presupposes the view that rationality is only a matter of internal coherence, which underlies the subjectivist to approach Bayesian statistics (Savage 1972). In contrast, my preferred epistemology, being achievabilist convergentism, is designed to capture the practice of frequentist statistics (as explained in section \ref{sec-freq-testing}). The debate between Bayesian statistics and frequentist statistics is a long-standing battle fought along an extensive frontline that spans philosophy and science (Lin 2024$b$). So, it would be immature to arbitrate between van Fraassen's explanation and mine given the limited space of this article.

Yet the contrast between van Fraassen's and my explanations of the irreconcilability highlights an important point---a point regarding the methodology of doing epistemology of science. Science is continuous with epistemology, and Quine (1969) takes a junction point to be psychology. Yet I believe that he missed another junction point, statistics and machine learning, as those areas of science are already richly fused with various epistemological thoughts, explicitly or implicitly. If we want to find a good epistemology of scientific inference to explain the irreconcilability of the scientific realism debate, philosophers should take those areas of science more seriously. Van Fraassen has offered an explanation aligning with the subjectivist approach to Bayesian statistics. Yet there are many other epistemological ideas in statistics and machine learning that need to be explored before we can determine which of the possible explanations for the irreconcilability is better, or whether some of them are complementary. I hope I have offered a step in this direction, providing an explanation aligned with the frequentist approach in statistics and learning theory in machine learning.



This brings me to the most important message I want to convey. Statistics and machine learning, largely overlooked in the realism debate thus far, represent scientists' deliberate efforts to develop ideas and tools for their own epistemological purposes. These epistemological endeavors deserve greater understanding to advance our deeply challenging epistemological debates, even if only by a small step. This is how I arrived at my proposal: achievabilist convergentism, a reconstruction of a significant part of statistics and machine learning. It provides a common ground on which both realists and anti-realists can stand, thriving in their own ways (as argued in sections \ref{sec-instrumentalism} and \ref{sec-realism}, respectively). Furthermore, this common ground helps isolate a particularly deep, distinctively epistemic root of the irreconcilability in the realism debate (as argued in section \ref{sec-deepest}). Even if my proposal of achievabilist convergentism ultimately proves mistaken, my broader suggestion remains: epistemologists in philosophy of science have much to explore and learn from the practice of statistics and machine learning.



\section*{Acknowledgements} I am grateful to Konstantin Genin, Jun Otsuka, Eddy Chen, and I-Sen Chen for their insightful discussions. I am especially indebted to Kevin Kelly and Elliott Sober, who have profoundly shaped my perspectives on scientific realism and anti-realism, respectively.

\appendix

\section*{Appendix: Proof}

The following proves the claim that, {\em in Perrin's case, any method violating Ockham's realist razor violates either (1) almost everywhere convergence, (2) maximal domain, or (3) stability}. Recall that Ockham's realist razor requires the following:
	\op 
	\im {\sc (Long-Run Realist)} Converge to the truth at every simple world, where atomic theory is true (thereby sacrificing convergence at their empirically equivalent counterparts).
	\im {\sc (Short-Run Ockham)} Never, ever, infer the complex hypothesis (the negation of atomic theory) whenever the evidence is still compatible with the simple one.
	\im {\sc (Deduction)} Whenever the evidence is incompatible with the simple hypothesis, infer (deductively) the complex hypothesis, which is the negation of atomic theory.
	\ed 
The last requirement, Deduction, can be written into the definition of inference methods without any harm. So, we only need to discuss violation of the first two conditions. 

There are only three ways to violate the Long-Run Realist condition.

\begin{figure}[ht]
	\centering \includegraphics[width=.7\textwidth]{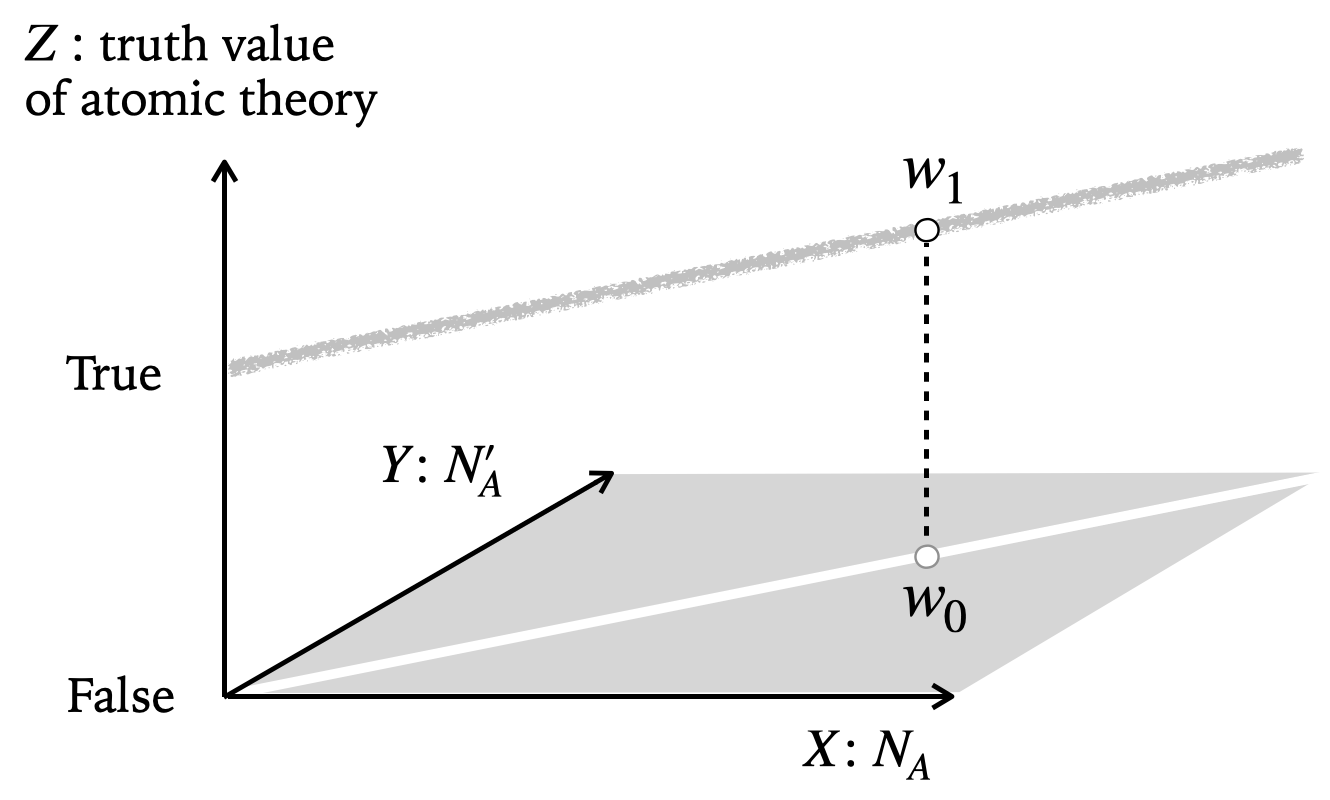}
	\caption{First way of violating the Long-Run Realist condition}
	\label{fig-way1}
	\end{figure}

{\em Way 1:} the domain of convergence misses only one point $w_1$ on the angel hair strand, and also misses its empirically equivalent counterpart $w_0$ on the lasagna sheet (figure \ref{fig-way1}). 
In this case, the criterion of maximal domain is violated, for we can design an alternative method whose domain of convergence extends the present one by adding either world $w_0$ or $w_1$.

\begin{figure}[ht]
	\centering \includegraphics[width=.7\textwidth]{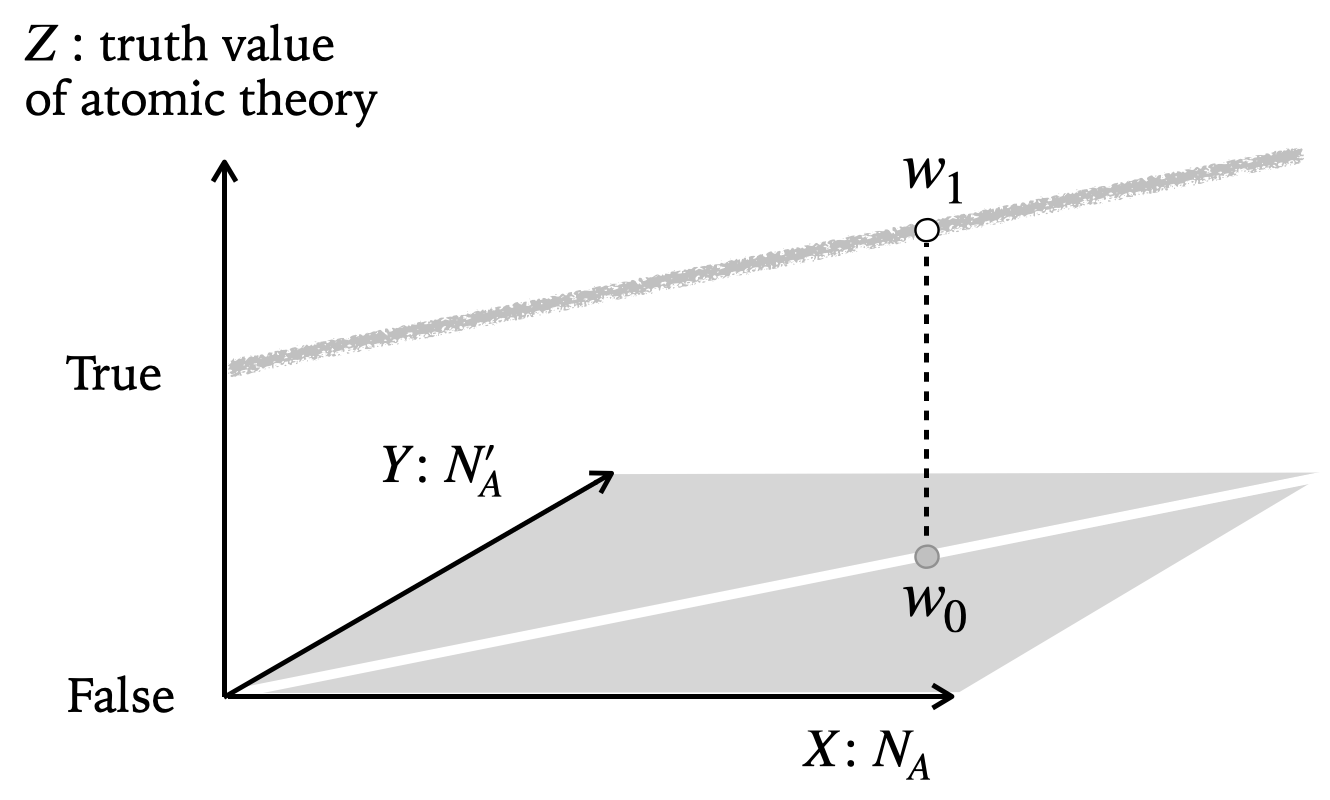}
	\caption{Second way of violating the Long-Run Realist condition}
	\label{fig-way2}
	\end{figure}

{\em Way 2:} the domain of convergence misses only one point $w_1$ on the angel hair strand, but covers its empirically equivalent counterpart $w_0$ on the lasagna sheet (figure \ref{fig-way2}). 
In this case, the criterion of stability is violated. To see why, note that this method must infer the negation of atomic theory given some evidence $E$ as a rectangular prism containing $w_0$. Now, suppose for {\em reductio} that this method adheres to stability. Then, by stability, this method must still infer the negation of atomic theory given any evidence included in the rectangular prism $E$. It follows that this method converges to the truth in all complex (atomless) worlds in $E$, and thus fails to converge to the truth in all simple worlds in $E$, which is an interval rather than a single point $w_1$ on the angel hair strand---contradiction. So, the present inference method must violate stability.

\begin{figure}[ht]
	\centering \includegraphics[width=.7\textwidth]{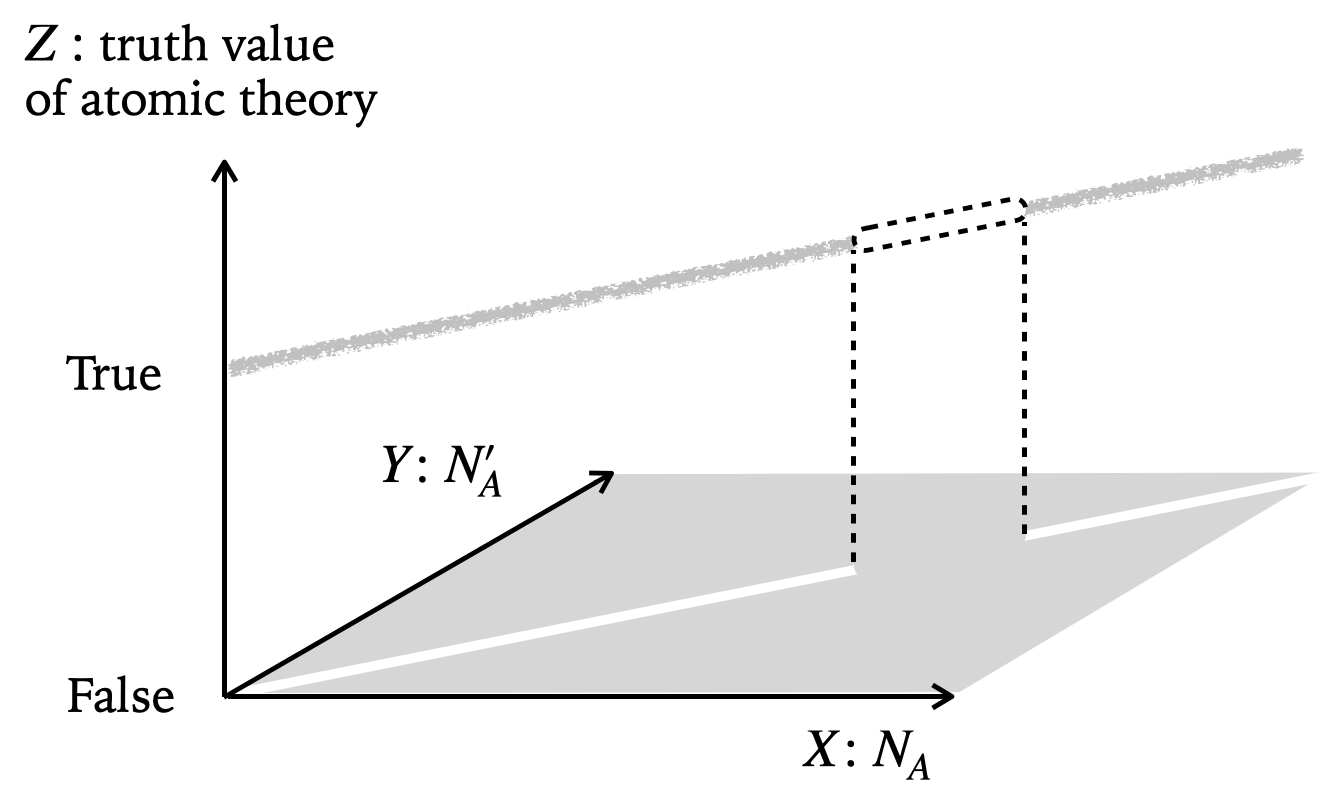}
	\caption{Third way of violating the Long-Run Realist condition}
	\label{fig-way3}
	\end{figure}

{\em Way 3:} the domain of convergence misses at least an interval on the angel hair strand (figure \ref{fig-way3}). 
In this case, the domain of convergence fails to come arbitrarily close to some points in that interval. The criterion of almost everywhere convergence is thus violated, by the Denseness Principle.

These cover all possible ways of violating the Long-Run Realist condition. 

Now, consider an inference method that violates the Short-Run Ockham condition. So, it infers the complex hypothesis when the evidence $E$ is still compatible with the simple hypothesis. Then, there are only two possibilities. Either this method violates stability, or it does not. In the latter case, the adherence to stability requires this inference method to never retract the complex hypothesis whenever more evidence is acquired in addition to $E$, and thus the present case reduces to the case of figure \ref{fig-way3} as discussed above, violating almost everywhere convergence.

These are the only ways of violating Ockham's realist razor. In each of these, there is failure of almost everywhere convergence, maximal domain, or stability. Q.E.D.


\section*{References}

\begin{description}
\item Aho, K., Derryberry, D., \& Peterson, T. (2014). Model selection for ecologists: the worldviews of AIC and BIC. {\em Ecology, 95}(3), 631-636.

\item Achinstein, P. (2001). {\em Evidence for molecules: Jean Perrin and molecular reality}. Oxford University Press.

\item Akaike, H. (1973). Information theory and an extension of the maximum likelihood principle. In B. N. Petrov \& F. Csaki (Eds.), {\em Proceedings of the 2nd International Symposium on Information Theory} (pp. 267-281). Akademiai Kiado.

\item Arkhangel'skii, A. V., \& Fedorchuk, V. V. (1990). {\em The basic concepts and constructions of general topology}. Springer.

\item Austern, M., \& Zhou, W. (2020). Asymptotics of cross-validation. {\em arXiv preprint}. https://arxiv.org/abs/2020.02752

\item Bayle, P., Bayle, A., Janson, L., \& Mackey, L. (2020). Cross-validation confidence intervals for test error. {\em Advances in Neural Information Processing Systems, 33}, 16339-16350.

\item Burnham, K. P., Anderson, D. R., \& Huyvaert, K. P. (2011). AIC model selection and multimodel inference in behavioral ecology: some background, observations, and comparisons. {\em Behavioral ecology and sociobiology, 65}, 23-35.

\item Chakravartty, A. (2017). Scientific Realism. Zalta, E. N. (ed.) {\em The Stanford Encyclopedia of Philosophy (Summer 2017 Edition)}, \\URL = $<$\texttt{https://plato.stanford.edu/archives/sum2017/entries/scientific-realism/}$>$.

\item Chalmers, A. (2011). Drawing philosophical lessons from Perrin's experiments on Brownian motion: A response to van Fraassen. {\em Studies in History and Philosophy of Science Part A, 42}(1), 63-72.

\item Claeskens, G., \& Hjort, N. L. (2008). {\em Model selection and model averaging}. Cambridge University Press.

\item Comesa\~{n}a, J. \& Klein. P. (2024). Skepticism. Zalta, E. N. \& Nodelman, U. (ed.) {\em The Stanford Encyclopedia of Philosophy (Spring 2024 Edition)}, \\URL = $<$\texttt{https://plato.stanford.edu/archives/spr2024/entries/skepticism/}$>$.

\item Diaconis, P., \& Freedman, D. (1986). On the consistency of Bayes estimates. {\em The Annals of Statistics, 14}(1), 1-26.

\item Diestel, R., Leader, I., Propp, J., \& Simons, S. (2014). {\em The joys of Haar measure}. Cambridge University Press.

\item Ding, J., Tarokh, V., \& Yang, Y. (2018). Model selection techniques: An overview. {\em IEEE Signal Processing Magazine, 35}(6), 16-34.

\item Einstein, A. (1905). On the movement of small particles suspended in a stationary liquid demanded by the molecular-kinetic theory of heat. {\em Annalen der Physik, 17}, 549-560.

\item Einstein, A. (1907). Theoretical observations on the Brownian motion. {\em Annalen der Physik, 19}, 371-381.

\item Fisher, R. A. (1925). {\em Statistical Methods for Research Workers}, Oliver \& Boyd.

\item Fedorchuk, V. V. (1990). {\em The fundamentals of dimension theory}. Springer.

\item Fitzpatrick, S. (2013). Doing away with the no miracle argument. In V. Karakostas \& D. Dieks (Eds.), {\em EPSA11 perspectives and foundational problems in philosophy of science (The European Philosophy of Science Association Proceedings, Vol. 2)} (pp. 141-151). Springer.

\item Forster, M., \& Sober, E. (1994). How to tell when simpler, more unified, or less ad hoc theories will provide more accurate predictions. {\em The British Journal for the Philosophy of Science, 45}(1), 1-35.

\item Freedman, D. (1963). On the asymptotic behavior of Bayes' estimates in the discrete case. {\em The Annals of Mathematical Statistics, 34}(4), 1386-1403.

\item Genin, K. (2018). {\em The topology of statistical inquiry} (PhD Dissertation). Carnegie Mellon University.


\item Hudson, R. (2020). The reality of Jean Perrin's atoms and molecules. {\em The British Journal for the Philosophy of Science, 71}, 33-58.

\item Hunt, B. R., Sauer, T., \& Yorke, J. A. (1992). Prevalence: A translation-invariant ``almost every'' on infinite-dimensional spaces. {\em Bulletin of the American Mathematical Society, 27}(2), 217-238.

\item Kelly, K. T. (1996). {\em The logic of reliable inquiry}. Oxford University Press.

\item Kelly, K. T. (2011). Simplicity, truth, and probability. In Bandyopadhyay, P. S., \& Forster, M. R. (eds.) {\em Philosophy of statistics} (pp. 983-1024). North-Holland.

\item Kelly, K. T., Genin, K., \& Lin, H. (2016). Realism, rhetoric, and reliability. {\em Synthese, 193}, 1191-1223

\item Lin, H. (2019). The hard problem of theory choice: A case study on causal inference and its faithfulness assumption. {\em Philosophy of Science, 86}(5), 967-980.

\item Lin, H. (2022). Modes of convergence to the truth: Toward a better epistemology of induction. {\em Journal of Philosophical Logic, 51}(2), 277-310.

\item Lin, H. (2024$a$). Internalist reliabilism in statistics and machine learning: Thoughts on Jun Otsuka's thinking about statistics. {\em The Asian Journal of Philosophy 3}(2), 1-11.

\item Lin, H. (2024$b$). To be a frequentist or Bayesian? Five positions in a spectrum. \textit{Harvard Data Science Review}, 6(3), doi: 10.1162/99608f92.9a53b923

\item Lin, H. (forthcoming $a$). Convergence to the truth. In K. Sylvan, E. Sosa, J. Dancy \& M. Steup (Eds.), \textit{The Blackwell Companion to Epistemology}, 3rd Edition. Wiley Blackwell.

\item Lin, H. (forthcoming $b$). Frequentist statistics as internalist reliabilism. In Y. Shan (ed.), \textit{Integrating Philosophy of Science and Epistemology}. Synthese Library. Springer. 

\item Lin, H., \& Zhang, J. (2020). On learning causal structures from non-experimental data without any faithfulness assumption. {\em Proceedings of Machine Learning Research, 21}(1), 1-36.

\item Li, J. (2023). Asymptotics of K-fold cross-validation. {\em Journal of Artificial Intelligence Research, 78}, 491-526.

\item Lugosi, G., \& Zeger, K. (1995). Nonparametric estimation via empirical risk minimization. {\em IEEE Transactions on Information Theory, 41}(3), 677-687.

\item Lugosi, G., \& Zeger, K. (1996). Concept learning using complexity regularization. {\em IEEE Transactions on Information Theory, 42}(1), 48-54

\item Maddy, P. (1997). {\em Naturalism in mathematics}. Oxford University Press.

\item Magnus, P. D., \& Callender, C. (2004). Realist ennui and the base rate fallacy. {\em Philosophy of Science, 71}(3), 320-338.

\item Meek, C. (1995). Strong-completeness and faithfulness in Bayesian networks. In {\em Proceedings of the eleventh conference on uncertainty in artificial intelligence}, Morgan Kaufmann, pp. 411-418.

\item Murtaugh, P. (2014). In defense of p-values. {\em Ecology, 95}(3), 611-617.

\item Nishii, R. (1984). Asymptotic properties of criteria for selection of variables in multiple regression. {\em The Annals of Statistics, 12}(1), 758-765.

\item Oxtoby, J. C. (1980). {\em Measure and category}. Springer.

\item Peirce, C. S. (1902$a$). Reasoning. In Baldwin, J. M. (ed,) {\em Dictionary of Philosophy and Psychology}. New York: Macmillan. 

\item Peirce, C. S. (1902$b$). Validity. In Baldwin, J. M. (ed,) {\em Dictionary of Philosophy and Psychology}. New York: Macmillan. 


\item Perrin, J. (1910). {\em Brownian movement and molecular reality}. Annales de Chimie et de Physique.

\item Psillos, S. (1999). {\em Scientific realism: How science tracks truth}. Routledge.

\item Psillos, S. (2011). Moving molecules above the scientific horizon: On Perrin's case for realism. {\em Journal for General Philosophy of Science, 42}, 339-363.

\item Putnam, H. (1965). Trial and error predicates and the solution to a problem of Mostowski. {\em Journal of Symbolic Logic, 30}(1), 49-57.

\item Putnam, H. (1975). {\em Mathematics, matter, and method}. Cambridge University Press.

\item Quine, W. V. O. (1969). Epistemology naturalized. In \textit{Ontological relativity and other essays} (pp. 69-90). Columbia University Press.

\item Reichenbach, H. (1938). {\em Experience and prediction: An analysis of the foundation and the structure of knowledge}. University of Chicago Press.

\item Rowbottom, D. P. (2011). The instrumentalist's new clothes. {\em Philosophy of Science, 78}(5): 1200-1211.

\item Saatsi, J. (2017). Replacing recipe realism. {\em Synthese, 194}(9), 3233-3244.

\item Savage, L. J. (1972). {\em The foundations of statistics}, 2nd edition. Dover Publications, Inc.


\item Schulte, P. (1999). {\em Means-ends epistemology}. {\em The British Journal for the Philosophy of Science, 50}(1), 1-31.


\item Shalev-Shwartz, S., \& Ben-David, S. (2014). {\em Understanding machine learning: From theory to algorithms}. Cambridge University Press.

\item Shibata, R. (1981). An optimal selection of regression variables. {\em Biometrika, 68}(1), 45-54.

\item Shao, J. (1997). An asymptotic theory for linear model selection. {\em Statistica Sinica, 7}(1), 221-242.

\item Shao, J. (2003). {\em Mathematical statistics}. Springer.

\item Sober, E. (2002). Instrumentalism, parsimony, and the Akaike framework. {\em Philosophy of Science, 69}(S3), S112-S123.

\item Spirtes, P., Glymour, C., \& Scheines, R. (2001). {\em Causation, prediction, and search}. MIT press.

\item Stanford, P. K. (2016). Instrumentalism: Global, local, and scientific. In P. Humphreys (Ed.), {\em The Oxford Handbook of Philosophy of Science}. Oxford University Press, 319-336.

\item Schwarz, G. (1978). Estimating the dimension of a model. {\em The annals of statistics 6}(2): 461-464.

\item Vapnik, V. N. (2000). {\em The nature of statistical learning theory} (2nd ed.). Springer.

\item Vapnik, V. N., \& Chervonenkis, A. Y. (1971). On the uniform convergence of relative frequencies of events to their probabilities. {\em Theory of Probability \& Its Applications, 16}(2), 264-280.

\item van Fraassen, B. C. (1989). {\em The scientific image}. Oxford University Press.

\item van Fraassen, B. C. (1989). {\em Laws and symmetry}. Oxford University Press.

\item van Fraassen, B. C. (1994). Gideon Rosen on constructive empiricism. {\em Philosophical Studies, 74}(2), 179-192.

\item van Fraassen, B. C. (2009). The perils of Perrin, in the hands of philosophers. {\em Philosophy of Science, 143}, 5-24.
\end{description}

\end{document}